\documentclass[12pt]{article}
\pdfoutput=1
\usepackage[utf8x]{inputenc}
\usepackage[colorlinks,linkcolor=blue,citecolor=blue,bookmarks,bookmarksnumbered]{hyperref}
\usepackage[scaled=0.85]{helvet}
\usepackage{amsmath,amssymb,accents,mathrsfs,XoohmE}
\usepackage{graphicx,xcolor}
\usepackage{booktabs}
\usepackage{multirow}
\usepackage{placeins}
\usepackage{amsmath}

\usepackage{enumitem}

\usepackage{XoohmE}

\let\TC=\textcolor
\definecolor{Hey}{rgb}{.9,.05,.4}
\definecolor{orange}{rgb}{1,.5,0}
\definecolor{plum}{rgb}{.4,0,.6}
\definecolor{R}{rgb}{1,0,0}
\definecolor{G}{rgb}{0,1,0}
\definecolor{B}{rgb}{0,0,1}

\long\def\CMTred#1{\leavevmode\TC{red}{\sf#1}}

\long\def\CMTR#1{\leavevmode\TC{R}{\sf#1}}

\long\def\CMTB#1{\leavevmode\TC{B}{\sf#1}}

\def\pa{\partial}        
\def\Tilde#1{\widetilde{#1}}                    
\def\Hat#1{\widehat{#1}}                        
\def\Bar#1{\overline{#1}}                       
\def\leftrightarrowfill{$\mathsurround=0pt \mathord\leftarrow \mkern-6mu
        \cleaders\hbox{$\mkern-2mu \mathord- \mkern-2mu$}\hfill
        \mkern-6mu \mathord\rightarrow$}
\def\dvec#1{\vbox{\ialign{##\crcr
        \leftrightarrowfill\crcr\noalign{\kern-1pt\nointerlineskip}
        $\hfil\displaystyle{#1}\hfil$\crcr}}}           
\def\dt#1{{\buildrel {\hbox{\LARGE .}} \over {#1}}}     

\def\fracm#1#2{\hbox{\large{${\frac{{#1}}{{#2}}}$}}}
\def\frac#1#2{{\textstyle{#1\over\vphantom2\smash{\raise.20ex
        \hbox{$\scriptstyle{#2}$}}}}}                   
\def\sfrac#1#2{{\vphantom1\smash{\lower.5ex\hbox{\small$#1$}}\over
        \vphantom1\smash{\raise.4ex\hbox{\small$#2$}}}} 
\def\bfrac#1#2{{\vphantom1\smash{\lower.5ex\hbox{$#1$}}\over
        \vphantom1\smash{\raise.3ex\hbox{$#2$}}}}       
\def\afrac#1#2{{\vphantom1\smash{\lower.5ex\hbox{$#1$}}\over#2}}    
\def\on#1#2{\mathop{\null#2}\limits^{#1}}               

\def\be{\begin{equation}}
\def\ee{\end{equation}}
\def\bea{\begin{eqnarray}}
\def\eea{\end{eqnarray}}

\def\dt#1{\on{\hbox{\bf .}}{#1}}                
\def\Dot#1{\dt{#1}}
\def\IR{\relax{\rm I\kern-.18em R}}
\def\binomial#1#2{\left(\,{\buildrel 
{\raise4pt\hbox{$\displaystyle{#1}$}}\over
{\raise-6pt\hbox{$\displaystyle{#2}$}}}\,\right)}

\def\[{\lfloor{\hskip 0.35pt}\!\!\!\lceil}
\def\]{\rfloor{\hskip 0.35pt}\!\!\!\rceil}

\usepackage[aligntableaux=center,boxsize=0.9em]{ytableau}

\newcommand{\singlet}{\ytableaushort{{\none[{\bm\cdot}]}}}

\definecolor{skyblue}{rgb}{0.12, 0.46, 1.00}
\definecolor{brightpink}{rgb}{1.0, 0.0, 0.5}
\definecolor{darkgreen}{rgb}{0.10, 0.75, 0.24}

\newcommand{\tinytwo}{{\raisebox{-0.03em}{\scriptsize $2$}}}
\newcommand{\tinytwobar}{{\raisebox{-0.03em}{\scriptsize $\overline{2}$}}}

\usepackage[T1]{pbsi}

\newcommand\mathbsi[1]{%
    \mathord{\text{\bsifamily #1}}%
}

\newcommand{\AG}{\mathbsi{G}}

\catcode`@=11
\def\un#1{\relax\ifmmode\@@underline#1\else
        $\@@underline{\hbox{#1}}$\relax\fi}
\catcode`@=12

\def\fracm#1#2{\hbox{\large{${\frac{{#1}}{{#2}}}$}}}

\def\ad{{\kern0.5pt
                   \alpha \kern-5.05pt
\raise5.8pt\hbox{$\textstyle.$}\kern
0.5pt}}

\def\Dot#1{{\kern0.5pt
     {#1} \kern-5.05pt \raise5.8pt\hbox{$\textstyle.$}\kern
0.5pt}}

\def\rD{{\rm D}}
\def\rI{{\rm I}}

\def\rL{{\rm L}}
\def\rR{{\rm R}}

\def\ca{{\cal A}}

\def\cf{{\cal F}}

\def\cv{{\cal V}}

\numberwithin{equation}{section}

\begin{document}

\thispagestyle{empty}
\noindent{\small
\hfill{$~~$}  \\ 
{}
}
\begin{center}
{\large \bf
Adynkra Genomes, Adynkrafields, and the  \\ 
4D, $\bm{\cal N}$ = 1 Supergravity
Superfield Prepotential \\
\vskip4pt
$~~$  
}  \\   [8mm]
{\large {
S.\ James Gates, Jr.\footnote{gatess@umd.edu}${}^{,a, b}$ and
Yangrui Hu\footnote{huyangrui@bimsa.cn}${}^{c, d}$
}}
\\*[6mm]
\emph{
\centering
$^{a}$University of Maryland, Department of Physics
\\[1pt]
J.\ S.\ Toll Hall, Room 1117,
College Park, MD 20742-4111, USA
\\[10pt]
$^{b}$University of Maryland, School of Public Policy
\\[1pt]
T. Marshall Hall,
College Park, MD 20742-5035, USA
\\[10pt]
$^{c}$Perimeter Institute for Theoretical Physics
\\[1pt]
Waterloo, ON N2L 2Y5, Canada
\\[10pt]
and
\\[10pt]
$^{d}$Beijing Institute of Mathematical Sciences and Applications 
\\[1pt]
Beijing, 101408, China
}
 \\*[50mm]
{ ABSTRACT}\\[05mm]
\parbox{142mm}{\parindent=2pc\indent\baselineskip=14pt plus1pt
A reimagining of the supergravity prepotential formulation of 4D,
$\cal N$ = 1 supergravity and its Salam-Strathdee superfield superconformal gauge
group is presented.
} \end{center}
\vfill
\vfill
\clearpage

\newpage
{\hypersetup{linkcolor=black}
\tableofcontents
}

\newpage
\section{Introduction}
\label{sec:INTRO}

$~~~~$ Many mathematical challenges remain in the quest for a comprehensive and logically complete
theory of supergravity in the context of 
Salam-Strathdee superspace.

The existence of a theory of supergravity emerged through the use of component-level formulations \cite{SG1,SG2}.  Though mostly overlooked, the comprehensive presentation in the first of these works was dependent on the then-available IT technology and adroit coding. A step toward the use of superfields, without actually doing
so \cite{BrtnL}, was undertaken by embedding the graviton, gravitino, and axial vector chirality
gauge fields into the components of a vector supermultiplet.  Soon afterward, 
investigations in superspace started.  This activity can be divided into three topical foci.

Within the domain of superfields, a first wave of exploration took place in a series  \cite{SFSGx1,SFSGx2,OS1,SFSGwg1,SFSGwg2} of papers.  However, it is notable that these investigations did {\em {not}} include consideration of superspace supergeometry.  Instead, these studies focused on the prepotential superfield that was
needed to serve as the foundation of the component formulations that emerged in the initial research investigations.

There arose a second wave of activity in this domain initiated
by the works of \cite{SFSGwz1,SFSGwz2}.  The focus of these works
was almost exclusively upon superspace supergeometrical aspects 
related to supergravity.  These works even uncovered the interesting
fact that the dynamics of superspace supergravity can be investigated
{\em {without}} knowledge of an underlying action that depends on
the supergravity prepotential.  This is achieved by imposing constraints
on the supertensors that describe the supergeometry and their consistency
with a set of superspace Bianchi identities.

Finally, in a third wave \cite{SFSGwz3,SFSG,OS2,OS3}, the two distinct approaches were unified.  The superspace covariant derivative, subject to the constraints, was initially 
described in the second of these citations and completely solved in terms of prepotentials.  This gave rise to explicit solutions of the supertensors describing the supergeometry
in terms of the prepotentials.  In these papers, a unified and
comprehensive description including both the prepotentials and the curved
superspace geometry is presented in a transparent manner.

All of this activity occurred prior to the prominent rise of superstring theory.  
Since then, work has been completed in superspaces related to eleven and ten
dimensions \cite{11Dssp,10TAssp,10TBssp,10ssp1,10sssp2}.  Almost all of these works have a 
character that aligns with the second topical wave investigating the initial theory of supergravity in superspace.  To our knowledge, until recently, only a single work of the third topical character existed within the domain of these higher-dimensional superspaces.  
It can be seen in a solitary work by Howe, Nicolai, and van Proeyen \cite{HNVP} that
was written in 1982 about one of the three ten-dimensional superspaces.

Given this recitation, the reader may ask, ``What is the equivalent of the third topical
wave described above as it relates to {\em {all}} of the  eleven- and ten-dimensional superspaces?"

Unfortunately, for the eleven-dimensional, the IIA, and the IIB superspaces, the answer is a disappointing vacuum...until recently.  As these are the theories that arise precisely in the limits of M-Theory, F-Theory, and superstring theories, filling this hole in the research literature 
may offer new opportunities to deepen our understanding of this enterprise.
It has long been our viewpoint \cite{ENUF} that {\em {the}} major obstruction to progress in this area is the concept of the traditional Salam-Strathdee superfield itself.  In these high dimensions, Salam-Strathdee superfields are {\em {extraordinarily}} opaque in their ability 
to render transparent results at the level of component fields.

By the time of the declaration in \cite{ENUF}, we had already begun a program to ``deconstruct'' \cite{GRana1,GRana2}
Salam-Strathdee superfields in order
to investigate the sources of their extraordinary capacity to provide a firm foundation for the representation theory of supersymmetry.  Eventually, this led to the introduction of adinkras \cite{adnk1}.  Subsequently, their
existence indicated pathways connecting the representation theory of SUSY to graph and
network theory, error-correction and coding theory \cite{codes1,codes2,codes3}, Coxeter groups \cite{permutadnk}, algebraic topology, and Grothendieck's {\it {dessin d'enfant}} \cite{adnkGEO1,adnkGEO2}, permutahedra  \cite{PrmHDRA}.

An upgrade of the concept of adinkras to a
version of ``adinkra 2.0" occurred in 2020 with the introduction of ``adynkras."  This word is a portmanteau created
 by the combination of two words: ``adinkra" and ``Dynkin.'' Adynkras \footnote{By a fortunate accident, the substitution of the letter ``y" for the
letter ``i" in adinkra can also stand for the first letter in YT.} are constructed by replacing the nodes of adinkras with Young Tableaux (YT). It is critical to note that Dynkin Labels (DLs) and Young tableaux each possess multiplication rules. 
The basis for such multiplications is the
Littlewood-Richardson rule \cite{littlewoodrichardson} for YT.
The basis for the multiplication of DLs is the use of Klimyk's
formula \cite{KLMK} for the decomposition of products of DLs into sums of
DLs.  Thus, conceptually there are the following close correspondences
\newline
$$
\begin{array}{ccc}
{\rm {Young~Tableaux }} & {\longleftrightarrow} & {\rm {Dynkin~ Labels ~~}}\\
{\rm {Littlewood-Richardson~ Rule}} & {\longleftrightarrow} &{~~~~~} {\rm {Klimyk's ~
Formula}} {~~~~~~}\\
\end{array}
$$
\newline \noindent
that exist.
For our purposes, YT and DLs are used interchangeably.\,\footnote{Note that standard Young Tableaux (YTs) correspond to irreducible representations of symmetric groups and ${\rm SU}(N)$ groups, but \textit{not} of ${\rm SO}(N)$ groups. To extend the YT formalism to ${\rm SO}(N)$ groups, so that the index structure remains manifest, we introduced bosonic (blue-colored) and spinorial (red-colored) YTs \cite{Cntg-10D,Cntg-method,Cntg-lowD}. The irreducible bosonic and spinorial YTs satisfy additional constraints, which we refer to as irreducibility conditions, and which can be systematically derived from branching rules. A systematic setup for these conditions is developed in \cite{Cntg-method}, where we also introduced graphical rules, called ``tying rules", to implement them in certain cases.}
In fact, we illustrate adynkras with both types of nodes.  A useful attribute of adynkras is that they correspond to drawable graphs, and due to the relation between YT, DLs, the
Littlewood-Richardson rule, and Klimyk's formula, these graphs inherit a set of multiplication rules.

After the completion of adynkras, one is led to the concept of ``adynkrafields."
Adynkrafields are a graphical technology for supersymmetric field theories, just as adinkras were introduced as a graphical technology for the study of representations of SUSY theories.
The most important feature of adinkrafields is their scalable component-level transparency as the algorithmic architecture of the codes describing them is largely independent
of the spacetime dimension under study.  We turned this to our advantage recently in the work
\cite{Cntg-11D}, in which the first off-shell prepotential proposal for the embedding of 11D supergravity
was presented.  If the recent proposal succeeds, it will have established a ``beachhead"
for the 11D supergravity theory that is comparable to that created in the works \cite{SFSGx1,SFSGx2,OS1,SFSGwg1,SFSGwg2} for the 4D supergravity theory.

The 4D, $\cal N $ = 1 adynkra construction relies on the introduction of two distinct types of YT.  One 
type is associated with the representation theory of bosonic representations of the spin group.  These 
are typically constructed from ``blue boxes'' $\CMTB{\ydiagram{1}}$.  The other type of tableau is 
associated with the fermionic representations of the spin group.  These are typically constructed from 
``red boxes'' $\CMTred{\ydiagram{1}}$.  There are also mixed tableaux constructed from both blue and 
red boxes.  However, only a single red box is permitted in such mixed tableaux.  The reason for this, 
as explained in the work of \cite{Cntg-method}, is that any YT with an even number of red boxes can
generally be decomposed into a sum of blue boxes.  This is the graphical analog of Fierz identities.
This approach has one great advantage.

Since the information on the spins of the various particles in these constructions is accessed via
either Dynkin Labels or Young Tableaux, no explicit matrices are required.  In fact, our adynkra 
construction is basis-independent and works without using matrices.  The graphical versions
of the Fierz identities are thus independent of matrices.  This is a tremendous computational 
advantage in the domains where we use them.  The disadvantage is that these methods do not
naturally provide Clebsch-Gordan coefficients.
 
 To simulate the Grassmann coordinates of superspace, a ``wedge product'' must be introduced among 
 the red boxes in the
 calculation of an ``adynkra genome''. A genome here may be regarded as a level-by-level list of representations associated with a corresponding adynkra.  This is accomplished by incorporating the mathematical concept of
 ``plethysm'' \cite{Plethysm} into the application of Klimyk's formula when the wedge product is utilized for the construction related to an adynkra or its genome (see chapter \ref{sec:AGOvrVw}).  Finally, as there are two distinct types of ``red boxes," there
 are two distinct ``level parameters"
 denoted by $\ell$ and
 $\bar \ell$. 
 
 There is another feature of these level parameters that has not received comment
 in previous work.  We postulate that nilpotency conditions must be among their properties.  The reason
 for this is our demand that adynkrafields should mirror the properties of the usual $\theta$-coordinates of
 superspace.  So the order of the power at which the associated $\theta$-coordinates vanish determines the order of nilpotency for the level parameters.
 The same order of nilpotency can be determined without reference to $\theta$-coordinates by examining the lowest power of wedge products of red boxes at which the singlet representation 
appears.  The order of the nilpotency using this comparison is one power greater than the
power of the wedge product.

The construction of adynkras, like the original construction of the theory of 
supergravity, relies on the use of adroit coding algorithms and currently available IT technology.  The mathematical concepts described above are included in codes
that are accessible via the works of \cite{Cntg-11D,Cntg-10D,Cntg-method,Cntg-lowD}.

In these same works, we have been developing and curating this {\em {new}} tool based on a paradigm shift to fill 
this nearly empty research space of high-dimensional superspace supergeometry
{\em {and}} supergravity prepotentials.

Previously we applied this to investigations of Weyl covariance and superconformal prepotentials in 10D superspaces \cite{Cntg-Weyl}. We have also used adynkras to achieve, to our knowledge, the first complete
decomposition into component fields starting from eleven-dimensional superfields.
In particular, in \cite{Cntg-11D}, we stated that whether the introduction of adynkras
will be seen as successful ``will be settled by research in the future.''

As our work is boundary-breaking, it is valid to criticize\footnote{We thank one of the referees for bringing this
point to our attention.} the fact that the distinction between Young diagrams and Young tableaux is not consistently
maintained. The figures presented correspond to Young diagrams in the standard mathematical sense, rather
than tableaux.

The graphical objects appearing in our equations are used primarily to encode the shapes, index symmetries, and spin-group representations associated with the relevant Dynkin Labels. Consequently, when no entries are placed in the boxes, these objects may more precisely be called Young diagrams.

Our use of the term ``Young tableau,'' and the abbreviation YT follows a convention frequently adopted in the representation-theory and high-energy-physics literature, in which the terms ``Young diagram'' and ``Young tableau'' are sometimes used interchangeably when only the representation-defining box configuration is required. The calculations in the paper depend upon the associated representation, tensor symmetry, Dynkin Label, and product rules, and not upon a numerical filling of the boxes. Therefore, this terminological distinction does not alter any of the decompositions or equations presented.

To avoid ambiguity, we propose to add a clarification stating that, unless a filling of the boxes is explicitly introduced, the graphical objects denoted by YT in the paper should more precisely be understood as Young diagrams---or, in cases involving the bosonic and spinorial box conventions, as generalized or colored Young diagrams. We will reserve the term ``Young tableau'' in its strict sense for a diagram equipped with entries.

One nuance may be worth preserving: the paper distinguishes bosonic, spinorial, and mixed objects by different kinds or colors of boxes. These markings carry representation-theoretic information, but they are not ordinarily a tableau ``filling'' in the standard combinatorial sense. Therefore, ``colored Young diagrams'' or ``generalized Young diagrams'' would be particularly defensible terminology.

The purpose of this work is to further the investigation along this path.  We propose to do this by benchmarking these new concepts and tools against the standard formulation of 4D, $\cal N$ = 1 supergravity by comparing Salam-Strathdee superfields with adynkrafields.  We wish to elucidate
similarities and differences that should be expected to occur.

This work is organized in the following manner.

The second chapter reviews the superfield formulation of some familiar supermultiplets.  The first supermultiplet discussed is the prepotential for the supergravity supermultiplet.  A review is given of the projection method
(``borrowing" from the discussion in ``{\it {Superspace}}'' \cite{superspace}) to obtain the component field spectrum.
This is followed by a ``translation" of the component field spectrum into equivalent spectra in terms of DLs and YT.  Next, the same treatment is carried out for a spinor superfield because it is the gauge parameter for the supergravity superfield prepotential. 

The third chapter begins with a discussion of the relationship of adinkras to adynkras.
The importance of introducing level parameters is noted because this is an important aspect of how such formulations remain consistent while sometimes containing multiple copies of the
same YT at the same level in an adynkra.  The concept of an ``adynkra genome", a mathematical structure that is obtained by the exponentiation of level parameters multiplied by YT,
is presented.  As in the case of the adinkras, an adynkra genome may also be regarded
as a type of network.  Finally, the chapter is concluded by providing six such adynkra genomes as examples of how the formalism works.

The fourth chapter is devoted to passing from adynkra genomes to adynkrafields.  An adynkra
genome may be regarded as a projection operator, expressed in terms of YT, that picks out the
fields of a supermultiplet from among the space of all possible component fields.  Thus, an adynkrafield
results when a certain ``overlap'' is taken between the space of component fields and an
adynkra genome.  
$~~$ \newline

\section{\texorpdfstring{4D, ${\cal N}=1$ Superfields Discussions}{4D, N=1 Superfields Discussions}}
\label{sec:SFrv}

\subsection{\texorpdfstring{4D, ${\cal N}=1$ Axial Vector Superfield}{4D, N=1 Axial Vector Superfield}}
\label{sec:SFrv1}

In 4D, $\cal N$ = 1 superspace, a superfield with a vector index, $H_{\un{a}}$, contains components
that can be defined by the equations (below we use the conventions in {\it {Superspace}} \cite{superspace}),
\begin{equation}
\begin{gathered}
    h_{\un{a}} ~=~ H_{\un{a}} | ~~~, \\
    h_{\a\b\Dot{\b}} ~=~ \rD_{\b} H_{\a\Dot{\b}} | ~~~, \\
    h^{(2)}{}_{\un{a}} ~=~ \rD^{2} H_{\un{a}} | ~~~,~~~
    h_{\un{a}\un{b}} ~=~ -~ \fracm{1}{2} \, [ \rD_{\a} , \Bar{\rD}_{\Dot{\a}} ] H_{\b\Dot{\b}} | ~~~, \\
    \Bar{\psi}_{\un{a} \, \Dot{\b}} ~=~ -~ i \, \rD^{2} \Bar{\rD}_{\Dot{\a}} H_{\a\Dot{\b}} | ~~~, \\
    A_{\un{a}} ~=~ -~ \fracm{2}{3} \, \rD^{\b} \Bar{\rD}^{2} \rD_{\b} H_{\un{a}} | ~-~ 
    \fracm{1}{6} \, \e_{\un{a}\un{b}\un{c}\un{d}} \, \pa^{\un{b}} \, [ \rD^{\g} , \Bar{\rD}^{\Dot{\g}} ] H^{\un{d}} | ~~~,
\end{gathered}
\label{eqn:HaCompSS}
\end{equation}
along with the conjugate fields $\Bar{h}_{\Dot{\a}\b\Dot{\b}}$ , $\Bar{h}^{(2)}{}_{\un{a}}$ and $\psi_{\un{a} \, \b}$.

Apart from the analytic approach of writing out the $\theta$-expansion of the superfield and acting with a number of supercovariant derivatives on it, one can deduce the structure of the component field expansion using group-theoretic approaches developed recently \cite{Cntg-11D,Cntg-10D,Cntg-method,Cntg-lowD}. Especially
from the work in \cite{Cntg-lowD}, we know that a scalar superfield in 4D has components expressed in $\mathfrak{so}(4)$ irreducible representations as follows:
\begin{equation}
    \cv ~=~
    \begin{cases}
    {\rm level-0:}~~~\CMTB{[0,0]}~~,~~\\
    {\rm level-1:}~~~\CMTred{[1,0]} \oplus \CMTred{[0,1]} ~~,~~\\
    {\rm level-2:}~~~(2)\CMTB{[0,0]}\oplus\CMTB{[1,1]}~~,~~ \\
    {\rm level-3:}~~~ \CMTred{[1,0]} \oplus \CMTred{[0,1]} ~~,\\
    {\rm level-4:}~~~\CMTB{[0,0]}~~,
    \end{cases}
\label{eqn:VCompDyn}
\end{equation}
where the basic bosonic and spinorial Dynkin Labels correspond to index structures as
\begin{equation}
    \un{a} ~\equiv~ \CMTB{[1,1]} ~~~,~~~ 
    \a ~\equiv~ \CMTred{[1,0]} ~~~,~~~
    \Dot{\a} ~\equiv~ \CMTred{[0,1]} ~~~.
\label{eqn:DynIndex}
\end{equation}
This makes sense as $\un{a} = \a\Dot{\a}$, i.e.
\begin{equation}
    \CMTB{[1,1]} ~=~ \CMTred{[1,0]} ~\otimes~ \CMTred{[0,1]} ~~~.
    \label{equ:a=aad}
\end{equation}
This will be further elaborated below via Young Tableau notation.

The component content of $H_{\un{a}}$ is thus $\cv \otimes \CMTB{[1,1]}$. Explicitly, we have
\begin{equation}
    H_{\un a} ~=~
    \begin{cases}
    {\rm level-0:}~~~\CMTB{[1,1]}~~,~~\\
    {\rm level-1:}~~~\CMTred{[1,0]} \oplus \CMTred{[0,1]} \oplus \CMTred{[1,2]}\oplus \CMTred{[2,1]}~~,~~\\
    {\rm level-2:}~~~\CMTB{[0,0]}\oplus\CMTB{[2,0]}\oplus\CMTB{[0,2]}\oplus(2)\CMTB{[1,1]}\oplus\CMTB{[2,2]}~~,~~ \\
    {\rm level-3:}~~~ \CMTred{[1,0]} \oplus \CMTred{[0,1]} \oplus \CMTred{[1,2]}\oplus \CMTred{[2,1]}~~,\\
    {\rm level-4:}~~~\CMTB{[1,1]}~~.
    \end{cases}
\label{eqn:HaCompDyn}
\end{equation}
To interpret these components, we will follow the techniques developed in the papers \cite{Cntg-method} and \cite{Cntg-lowD}, but adapted to 4D in order to translate
Dynkin Labels into Young Tableaux (YT), then to index notation. Next, we compare (\ref{eqn:HaCompSS}) and (\ref{eqn:HaCompDyn}).

The first step is to translate Dynkin Labels into YT notation.

As discussed in \cite{Cntg-lowD}, in 4D, ${\cal N}=1$ superspace, given a bosonic irrep with Dynkin Label $\CMTB{[a,b]}$, its corresponding irreducible 
bosonic Young Tableau is composed of $\min\{a,b\}$ columns of one box and $|b-a|/2$ columns of two 
vertical boxes. 
Therefore,
the fundamental building blocks of an irreducible BYT (bosonic YT) are\,\footnote{Note that $\CMTB{\ydiagram{1}} = {\CMTB{\ydiagram{1}}}_{{\rm IR}}$ as it is already irreducible without putting on any irreducible conditions. Thus we omit the IR subscript for neat presentation.}
\begin{equation}
\begin{gathered}
    {\CMTB{\ydiagram{1}}} ~\equiv~ \CMTB{[1,1]}  ~~~,~~~
    {\CMTB{\ydiagram{1,1}}}_{{\rm IR},+} ~\equiv~ \CMTB{[0,2]} ~~~,~~~
    {\CMTB{\ydiagram{1,1}}}_{{\rm IR},-} ~\equiv~ \CMTB{[2,0]} ~~~,
\end{gathered} \label{equ:BYTbasic_4D}
\end{equation}
where ``+'' indicates self-dual 2-form, and ``$-$'' indicates anti-self-dual 2-form. A generic 2-form without specific duality constraint could be written as
\begin{equation}
    {\CMTB{\ydiagram{1,1}}}_{{\rm IR}} ~\equiv~ {\CMTB{\ydiagram{1,1}}}_{{\rm IR},+} ~\oplus~ {\CMTB{\ydiagram{1,1}}}_{{\rm IR},-} 
    ~\equiv~ \CMTB{[0,2]} ~\oplus~ \CMTB{[2,0]} ~~~.
\end{equation}

For spinorial irreps, by adopting the Weyl convention, the basic SYTs (spinorial YTs) are
\begin{equation}
    \CMTred{\ytableaushort{\tinytwo}} ~\equiv~ \CMTred{[1,0]} ~~,~~ 
    \CMTred{\ytableaushort{\tinytwobar}} ~\equiv~ \CMTred{[0,1]} ~~~.
\label{equ:SYTbasic_4D}
\end{equation}
For a general spinorial irrep $\CMTred{[a,b]}$, one follows the convention in \cite{Cntg-method}. When $b>a$, one interprets it as $\CMTred{[0,1]}$ ``+'' some bosonic Dynkin Labels; when $a>b$, one interprets it as $\CMTred{[1,0]}$ ``+'' some bosonic Dynkin Labels. This ensures conjugation preserves dimensionality. An example that illustrates this well is
\begin{align}
    \CMTred{[1,2]} ~=&~ \CMTB{[1,1]} ~+~ \CMTred{[0,1]} ~=~ {\CMTB{\ydiagram{1}}\CMTred{\ytableaushort{\tinytwobar}}}_{\rm IR} ~~~, \\
    \CMTred{[2,1]} ~=&~ \CMTB{[1,1]} ~+~ \CMTred{[1,0]} ~=~ {\CMTB{\ydiagram{1}}\CMTred{\ytableaushort{\tinytwo}}}_{\rm IR} ~~~.
\end{align}

Therefore, the Young tableau description (which equivalently describes the index structures of corresponding component fields) of the component field content of a 4D, ${\cal N}=1$ vector superfield is as seen below
\begin{equation}
    H_{\un a} ~=~
    \begin{cases}
    {\rm level-0:}~~~{\CMTB{\ydiagram{1}}}~~,~~\\
    {\rm level-1:}~~~\CMTred{\ytableaushort{\tinytwo}} \oplus \CMTred{\ytableaushort{\tinytwobar}} \oplus {\CMTB{\ydiagram{1}}\CMTred{\ytableaushort{\tinytwobar}}}_{\rm IR}\oplus {\CMTB{\ydiagram{1}}\CMTred{\ytableaushort{\tinytwo}}}_{\rm IR}~~,~~\\
    {\rm level-2:}~~~\CMTB{\singlet}\oplus{\CMTB{\ydiagram{1,1}}}_{\rm IR,-}\oplus{\CMTB{\ydiagram{1,1}}}_{\rm IR,+}\oplus(2){\CMTB{\ydiagram{1}}}\oplus{\CMTB{\ydiagram{2}}}_{\rm IR}~~,~~ \\
    {\rm level-3:}~~~\CMTred{\ytableaushort{\tinytwo}} \oplus \CMTred{\ytableaushort{\tinytwobar}} \oplus {\CMTB{\ydiagram{1}}\CMTred{\ytableaushort{\tinytwobar}}}_{\rm IR}\oplus {\CMTB{\ydiagram{1}}\CMTred{\ytableaushort{\tinytwo}}}_{\rm IR}~~,\\
    {\rm level-4:}~~~{\CMTB{\ydiagram{1}}}~~.
    \end{cases}
\label{eqn:HaCompYT}
\end{equation}
To interpret all of these, let us utilize the techniques in \cite{Cntg-method}. This includes Littlewood's rule for BYT \cite{BYT-1,BYT-2,BYT-3,BYT-4,BYT-5,BYT-6,BYT-SYT} or its recent variant tying rule \cite{Cntg-method}, and tensor product rules between bosonic YTs and SYT developed by Murnaghan, Littlewood, King and others \cite{SYT-1,SYT-2,SYT-3,BYT-SYT}. The relevant tying rule is
\begin{equation}
    \CMTB{\ydiagram{2}} ~=~ {\CMTB{\ydiagram{2}}}_{\rm IR} ~\oplus~ \CMTB{\singlet} ~~~,
\end{equation}
while the relevant spinorial tensor product rules are
\begin{align}
    {\CMTB{\ydiagram{1}}} ~\otimes~ \CMTred{\ytableaushort{\tinytwo}} ~=&~~ {\CMTB{\ydiagram{1}}\CMTred{\ytableaushort{\tinytwo}}}_{\rm IR} ~\oplus~ \CMTred{\ytableaushort{\tinytwobar}} ~~~, \\
    {\CMTB{\ydiagram{1}}} ~\otimes~ \CMTred{\ytableaushort{\tinytwobar}} ~=&~~ {\CMTB{\ydiagram{1}}\CMTred{\ytableaushort{\tinytwobar}}}_{\rm IR} ~\oplus~ \CMTred{\ytableaushort{\tinytwo}} ~~~.
\end{align}
Therefore, one could rewrite (\ref{eqn:HaCompYT}) as
\begin{equation}
    H_{\un a} ~=~
    \begin{cases}
    {\rm level-0:}~~~ \CMTB{\ydiagram{1}} ~~~, \\
    {\rm level-1:}~~~ \CMTB{\ydiagram{1}} \otimes \CMTred{\ytableaushort{\tinytwo}} ~~\oplus~~ \CMTB{\ydiagram{1}} \otimes \CMTred{\ytableaushort{\tinytwobar}} ~~~, \\
    {\rm level-2:}~~~ (2) \, \CMTB{\ydiagram{1}} ~~\oplus~~ \CMTB{\ydiagram{1}} \otimes \CMTB{\ydiagram{1}} ~~~, \\
    {\rm level-3:}~~~ \CMTB{\ydiagram{1}} \otimes \CMTred{\ytableaushort{\tinytwobar}} ~~\oplus~~ \CMTB{\ydiagram{1}} \otimes \CMTred{\ytableaushort{\tinytwo}} ~~~, \\
    {\rm level-4:}~~~ \CMTB{\ydiagram{1}} ~~~.
    \end{cases}
\label{eqn:HaCompYTprod}
\end{equation}

The second step is to translate YT to index notation. The basic ones are
\begin{equation}
    \un{a} ~\equiv~ {\CMTB{\ydiagram{1}}} ~~~,~~~
    \a ~\equiv~ \CMTred{\ytableaushort{\tinytwo}} ~~~,~~~
    \Dot{\a} ~\equiv~ \CMTred{\ytableaushort{\tinytwobar}} ~~~,
\end{equation}
hence one obtains (\ref{eqn:DynIndex}). One could therefore do the final translation, 
\begin{equation}
    H_{\un a} ~=~
    \begin{cases}
    {\rm level-0:}~~~ h_{\un{a}} ~~~, \\
    {\rm level-1:}~~~ h_{\a\b\Dot{\b}} ~~~,~~~ \Bar{h}_{\Dot{\a}\b\Dot{\b}} ~~~, \\
    {\rm level-2:}~~~ h^{(2)}{}_{\un{a}} ~~~,~~~ \Bar{h}^{(2)}{}_{\un{a}} ~~~,~~~ h_{\un{a}\un{b}} ~~~, \\
    {\rm level-3:}~~~ \Bar{\psi}_{\un{a} \, \Dot{\b}} ~~~,~~~ \psi_{\un{a} \, \b} ~~~, \\
    {\rm level-4:}~~~ A_{\un{a}} ~~~.
    \end{cases}
\label{eqn:HaCompIndex}
\end{equation}
This conveys exactly the same content as (\ref{eqn:HaCompSS}). Note that some of these component fields are not irreducible. If one looks at (\ref{eqn:HaCompSS}), one would find that the analytic definitions do not impose any irreducible condition or symmetry on the indices. Therefore the whole story is consistent.

We end this section by noting that the
identification of the graviton as a
linear combination of YTs
\begin{equation}
h_{\un{a}\un{b}} ~\sim~ 
{\CMTB{\ydiagram{2}}}_{\rm IR} ~ \oplus ~
{\CMTB{\ydiagram{1,1}}}_{\rm IR} ~ \oplus ~ 
\CMTB{\singlet} ~=~
{\CMTB{\ydiagram{2}}}_{\rm IR} ~ \oplus ~
{\CMTB{\ydiagram{1,1}}}_{\rm IR,+} ~ \oplus ~
{\CMTB{\ydiagram{1,1}}}_{\rm IR,-} ~\oplus~
\CMTB{\singlet} ~~~,
\label{eqn:Hag}
\end{equation}
similarly for the gravitino 
\begin{equation}
\psi_{\un{a} \, \b} ~\sim~ {\CMTB{\ydiagram{1}}\CMTred{\ytableaushort{\tinytwo}}}_{\rm IR} ~\oplus~ \CMTred{\ytableaushort{\tinytwobar}}  ~~~,
\end{equation}
and finally for the axial vector auxiliary field
\begin{equation}
A_{\un a } ~\sim~ {\CMTB{\ydiagram{1}}} ~~~,
\end{equation}
that are located at three adjacent levels act as an adynkrafield ``genetic marker'' in four dimensions for how to identify superfields that are candidates for the supergravity prepotential.  With minor 
appropriate modifications, this statement applies to all dimensions and is the basis for the work seen in \cite{Cntg-Weyl}.

\subsection{\texorpdfstring{4D, ${\cal N}=1$ Spinor Superfields}{4D, N=1 Spinor Superfields}}

$~~~~$ Here we review the discussion of the gauge 
transformation of the supergravity prepotential superfield in Chapter Five of ``Superspace'' \cite{superspace}. The transformation of $H_{\un a}$ can be written 
in terms of the covariant derivatives of spinor superfields. 
\begin{equation}
    \d H_{\un{a}} ~=~ \rD_{\a} \Bar{L}_{\Dot{\a}} ~-~ \Bar{\rD}_{\Dot{\a}} L_{\a} ~~~,
\end{equation}
where
\begin{equation}
    \rD_{\a} ~=~ \pa_{\a} ~+~ i \, \fracm{1}{2} \, \Bar{\theta}^{\Dot{\a}} \pa_{\a\Dot{\a}} ~~~,~~~ 
    \Bar\rD_{\Dot\a} ~=~ \pa_{\Dot{\a}} ~+~ i \, \fracm{1}{2} \, {\theta}^{\a} \pa_{\a\Dot{\a}} ~~~.
\end{equation}

The component fields of $\Bar{\rD}_{\Dot{\a}} L_{\a}$ are defined by 
\begin{equation}
\begin{gathered}
    \x_{\un{a}} ~=~ \Bar{\rD}_{\Dot{\a}} L_{\a} | ~~~, \\
    L^{(1)}{}_{\a\b\Dot{\b}} ~=~ \rD_{\a} \Bar{\rD}_{\Dot{\b}} L_{\b} | ~~~,~~~ 
    \e_{\a} ~=~ \Bar{\rD}^{2} L_{\a} | ~~~, \\
    L^{(2)}{}_{\un{a}} ~=~ \rD^{2} \Bar{\rD}_{\Dot{\a}} L_{\a} | ~~~,~~~
    \s ~=~ \rD^{\a} \Bar{\rD}^{2} L_{\a} | ~~~,~~~
    \omega_{\a\b} ~=~ \fracm{1}{2} \, \rD_{(\a} \Bar{\rD}^{2} L_{\b)} | ~~~, \\
    \eta_{\a} ~=~ \rD^{2} \Bar{\rD}^{2} L_{\a} | ~~~,
\end{gathered}
\label{equ:4Dcomp_spinorSF}
\end{equation}
and similarly for the complex conjugates.

Finally, one obtains for the transformation law of the various component fields the expressions
\begin{equation}
\begin{split}
    \d h_{\un{a}} ~=&~ -~ 2 \, {\rm Re} \, \xi_{\un{a}} ~~~, \\
    \d h_{\a\b\Dot{\b}} ~=&~ ~ C_{\a\b} \, \Bar{\e}_{\Dot{\b}} ~-~ L^{(1)}{}_{\b\a\Dot{\b}} ~~~, \\
    \d \Bar{h}_{\Dot{\a}\b\Dot{\b}} ~=&~ -~ C_{\Dot{\a}\Dot{\b}} \, \e_{\b} ~-~ \Bar{L}^{(1)}{}_{\Dot{\b}\b\Dot{\a}} ~~~, \\
    \d h^{(2)}{}_{\un{b}} ~=&~ -~ L^{(2)}{}_{\un{b}} ~~~, \\
    \d \Bar{h}^{(2)}{}_{\un{b}} ~=&~ -~ \Bar{L}^{(2)}{}_{\un{b}} ~~~, \\
    \d h_{\un{a}\un{b}} ~=&~ -~ \big(\, C_{\Dot{\a}\Dot{\b}} \, \omega_{\a\b} ~-~ C_{\a\b} \, \Bar{\omega}_{\Dot{\a}\Dot{\b}} \,\big) ~+~ C_{\a\b} C_{\Dot{\a}\Dot{\b}} \, {\rm Re} \, \s ~+~ \pa_{\un{a}} \, {\rm Im} \, \xi_{\un{b}} ~~~, \\
    \d \psi_{\un{b} \a} ~=&~ ~ \pa_{\un{b}} \e_{\a} ~+~ i \, C_{\a\b} \, \Bar{\eta}_{\Dot{\b}} ~~~, \\
    \d \Bar{\psi}_{\un{b} \Dot{\a}} ~=&~ ~ \pa_{\un{b}} \Bar{\e}_{\Dot{\a}} ~+~ i \, C_{\Dot{\a}\Dot{\b}} \, \eta_{\b} ~~~, \\
    \d A_{\un{b}} ~=&~ -~ \fracm{2}{3} \, \pa_{\un{b}} \, {\rm Im} \, \s ~~~,
\end{split}
\label{equ:4Dcomp_laws}
\end{equation}
where $\un{a} = \a\Dot{\a}$ (see Eq.~\ref{equ:a=aad}) and $C_{\a\b}$, $C_{\Dot{\a}\Dot{\b}}$ are spinor metrics. 
It should be noted that these results define both the 4D, $\cal N$ = 1 
supergravity conformal fields and their transformations under the action
of the superconformal group.  The real part of $\xi{}_{\un a}$ can be used to set $h{}_{\un a}$ = 0.  The fermionic gauge parameter $L{}^{(1)}{}_{\b \a {\Dot \b}} $ can be used to 
set $h{}_{\a \, \b \Dot \b}  $ = 0. Finally, the bosonic gauge parameter $L{}^{(2)}{}_{\un a}$
can be used to set $ h{}^{(2)}{}_{\un a}$ = 0.  Thus, in a Wess-Zumino
gauge one is left with the component fields $h{}_{{\un a} {\un b}}$, 
${\psi}{}_{\un a}{}^{\b}$, and $A{}_{\un a}$ whose gauge transformation laws in this Wess-Zumino gauge are given by the last four lines of Eq.\ (\ref{equ:4Dcomp_laws}).

Consider the spinor superfield $L_{\a}$; its component content is actually ${\cal V}\otimes\CMTred{[0,1]}$\footnote{In the previous section we didn't talk about the positions of spinor indices because both scalar and vector superfields discussed above are actually real. In each level, component fields and their conjugate show up simultaneously. However, in this section, the position of the spinor index is nontrivial.} since the subscript $\a$ is equivalent to the conjugate of the superscript $\a$, i.e. superscript $\Dot\a$. Following the same conventions in \cite{Cntg-lowD}, the superscript $\a$ corresponds to $\CMTred{[1,0]}$ and the superscript $\Dot\a$ corresponds to $\CMTred{[0,1]}$. 

Explicitly, 
\begin{equation}
    L_{\a} ~=~
    \begin{cases}
    {\rm level-0:}~~~\CMTred{[0,1]}~~,~~\\
    {\rm level-1:}~~~\CMTB{[0,0]} \oplus \CMTB{[0,2]} \oplus \CMTB{[1,1]}~~,~~\\
    {\rm level-2:}~~~\CMTred{[1,0]}\oplus(2)\CMTred{[0,1]}\oplus\CMTred{[1,2]}~~,~~ \\
    {\rm level-3:}~~~ \CMTB{[0,0]} \oplus \CMTB{[0,2]} \oplus \CMTB{[1,1]}~~,\\
    {\rm level-4:}~~~\CMTred{[0,1]}~~.
    \end{cases}
\end{equation}

As we did in the vector superfield section, in order to understand the Dynkin Label content, first we translate Dynkin Labels to YT notation. 

\begin{equation}
    L_{\a} ~=~
    \begin{cases}
    {\rm level-0:}~~~\CMTred{\ytableaushort{\tinytwobar}}~~,~~\\
    {\rm level-1:}~~~\CMTB{\singlet} \oplus {\CMTB{\ydiagram{1,1}}}_{\rm IR,+} \oplus {\CMTB{\ydiagram{1}}}~~,~~\\
    {\rm level-2:}~~~(2)\CMTred{\ytableaushort{\tinytwobar}}\oplus \CMTB{\ydiagram{1}}\otimes\CMTred{\ytableaushort{\tinytwobar}}~~,~~ \\
    {\rm level-3:}~~~ \CMTB{\singlet} \oplus {\CMTB{\ydiagram{1,1}}}_{\rm IR,+} \oplus {\CMTB{\ydiagram{1}}}~~,\\
    {\rm level-4:}~~~\CMTred{\ytableaushort{\tinytwobar}}~~.
    \end{cases}
    \label{eqn:LalphaCompYT}
\end{equation}

Then we can translate YT to index notation. 
Note that the two-form $A_{\un a\un b}$ can be transformed into spinor notation by using spinor metrics (which are purely imaginary in our convention),
\begin{equation}
    A_{\un a\un b} ~=~ i\, C_{\Dot \a\Dot \b} A_{\a\b} ~+~ i\,C_{\a\b} A_{\Dot \a\Dot \b}  ~~~. 
    \label{eq:mp1}
\end{equation}
Define 
\begin{equation}
    A^{(+)}_{\un a\un b} ~=~ i\, C_{\Dot \a\Dot \b} A_{\a\b}  ~~~,~~~ 
    A^{(-)}_{\un a\un b} ~=~ i\,C_{\a\b} A_{\Dot \a\Dot \b}~~,
\label{eq:mp2}
\end{equation}
Equivalently, 
\begin{equation}
    A_{\a\b} ~=~ -i\,\fracm12\, C^{\Dot \a\Dot \b} A^{(+)}_{\un a\un b}   ~~~,~~~ 
    A_{\Dot \a\Dot \b} ~=~ -i\,\fracm12\,C^{\a\b} A^{(-)}_{\un a\un b}~~,
\label{eq:mp2a}
\end{equation}
and 
\begin{equation}
    A_{\un a\un b} ~=~ A^{(+)}_{\un a\un b} ~+~ A^{(-)}_{\un a\un b} ~~~. 
\label{eq:mp3}
\end{equation}
Here, because the spinor metrics satisfy $C_{\a\b} = - \, C_{\b\a}$ and $C_{\dot \a\dot \b} = - \, C_{\Dot \b\Dot \a} \,$, $A_{\a\b}$ and $A_{\Dot \a\Dot \b}$ have to be symmetric tensors and they each have three independent components. 
The SO(4) group tells us the same story. In terms of YTs, we have 
\begin{equation}
   {\CMTB{\ydiagram{1,1}}} ~=~  {\CMTB{\ydiagram{1,1}}}_{\rm IR,+} \oplus {\CMTB{\ydiagram{1,1}}}_{\rm IR,-} ~=~ \CMTB{\{\bar{3}\}} \oplus \CMTB{\{3\}} ~=~ \CMTB{[0,2]} \oplus \CMTB{[2,0]} ~~~. 
\end{equation}
Moreover, an application of Klimyk's formula reads
\begin{equation}
    \begin{split}
        \CMTred{[0,1]} ~\otimes~ \CMTred{[0,1]} ~=&~ \CMTB{[0,0]} ~\oplus~ \CMTB{[0,2]} ~~~,\\
        \CMTred{[1,0]} ~\otimes~ \CMTred{[1,0]} ~=&~ \CMTB{[0,0]} ~\oplus~ \CMTB{[2,0]} ~~~.
    \end{split}
\label{equ:22}
\end{equation}
Therefore, we can make the following identifications:
\begin{equation}
\begin{split}
    {\CMTB{\ydiagram{1,1}}}_{\rm IR,+} ~=&~ A^{(+)}_{\un a\un b}~{\rm or}~  A_{\a\b}~~,\\
    {\CMTB{\ydiagram{1,1}}}_{\rm IR,-} ~=&~  A^{(-)}_{\un a\un b}~{\rm or}~  A_{\Dot \a\Dot \b}~~.
\end{split}
\label{equ:translation-twoform}
\end{equation}
At this point, it is useful to discuss the multiplication of the red boxes
and tensor products needed to obtain results:
\begin{align}
    \CMTred{\ytableaushort{\tinytwo}}  ~\otimes~ \CMTred{\ytableaushort{\tinytwo}} ~=&~ \CMTB{\singlet} ~\oplus~ {\CMTB{\ydiagram{1,1}}}_{\rm IR,-} ~~~, \\
    \CMTred{\ytableaushort{\tinytwobar}}  ~\otimes~ \CMTred{\ytableaushort{\tinytwobar}} ~=&~ \CMTB{\singlet} ~\oplus~ {\CMTB{\ydiagram{1,1}}}_{\rm IR,+} ~~~, \\
    \CMTred{\ytableaushort{\tinytwo}}  ~\otimes~\CMTred{\ytableaushort{\tinytwobar}} ~=&~ \CMTB{\ydiagram{1}}_{\rm IR} ~~~, \\
    {\CMTB{\ydiagram{1}}} ~\otimes~ \CMTred{\ytableaushort{\tinytwo}} ~=&~ {\CMTB{\ydiagram{1}}\CMTred{\ytableaushort{\tinytwo}}}_{\rm IR} ~\oplus~ \CMTred{\ytableaushort{\tinytwobar}} ~~~, \\
    {\CMTB{\ydiagram{1}}} ~\otimes~ \CMTred{\ytableaushort{\tinytwobar}} ~=&~ {\CMTB{\ydiagram{1}}\CMTred{\ytableaushort{\tinytwobar}}}_{\rm IR} ~\oplus~ \CMTred{\ytableaushort{\tinytwo}} ~~~.
\end{align}

Relevant results for wedge product ${\bm \wedge}$ calculations include: 
first, we know
\begin{equation}
    \CMTred{\ytableaushort{\tinytwo}} ~{\bm \wedge}~ \CMTred{\ytableaushort{\tinytwo}} ~=~ \CMTB{\singlet}  ~~~,~~~
    \CMTred{\ytableaushort{\tinytwobar}} ~{\bm \wedge}~ \CMTred{\ytableaushort{\tinytwobar}} ~=~ \CMTB{\singlet}~~~.
\end{equation}
Next, consider the quadratic theta-monomial decompositions occurring in the scalar superfield. We can construct three different quadratic theta-monomials, $\theta^\a\theta^\b$, $\Bar{\theta}^{\Dot\a}\Bar{\theta}^{\Dot\b}$, and $\theta^\a\Bar{\theta}^{\Dot\b}$, which correspond to $\CMTred{\ytableaushort{\tinytwo}} ~{\bm \wedge}~ \CMTred{\ytableaushort{\tinytwo}}$, $\CMTred{\ytableaushort{\tinytwobar}} ~{\bm \wedge}~ \CMTred{\ytableaushort{\tinytwobar}}$, and $\CMTred{\ytableaushort{\tinytwo}} ~\otimes~ \CMTred{\ytableaushort{\tinytwobar}}$ respectively.
On the other hand, we can consider the quadratic theta-monomial decompositions as $[\,\theta^\a+\Bar{\theta}^{\Dot\a}\,]\,[\,\theta^\b+\Bar{\theta}^{\Dot\b}\,]$ and in YT language
\begin{equation}
    \begin{split}
        [\,\CMTred{\ytableaushort{\tinytwo}}\oplus \CMTred{\ytableaushort{\tinytwobar}} \,]~{\bm \wedge}~[\,\CMTred{\ytableaushort{\tinytwo}}\oplus \CMTred{\ytableaushort{\tinytwobar}} \,] ~=&~ \CMTred{\ytableaushort{\tinytwo}}~{\bm \wedge}~\CMTred{\ytableaushort{\tinytwo}} ~\oplus~\CMTred{\ytableaushort{\tinytwobar}}~{\bm \wedge}~\CMTred{\ytableaushort{\tinytwobar}} ~\oplus (2)\,\CMTred{\ytableaushort{\tinytwo}}~{\bm \wedge}~\CMTred{\ytableaushort{\tinytwobar}}~~~.
    \end{split}
\end{equation}
Therefore we obtain 
\begin{equation}
    \CMTred{\ytableaushort{\tinytwo}}~{\bm \wedge}~\CMTred{\ytableaushort{\tinytwobar}} ~=~ \CMTred{\ytableaushort{\tinytwo}} ~\otimes~ \CMTred{\ytableaushort{\tinytwobar}}~~~.
\end{equation}
In summary, the wedge product between the same two irreps/YTs is well defined and one can use the Plethysm function to carry out the calculation; the wedge product between two different irreps/YTs can be treated as the tensor product.

Gathering the results discussed above, the component content of $L_{\a}$ is:
\begin{equation}
    L_{\a} ~=~
    \begin{cases}
    {\rm level-0:}~~~l_{\a}~~,~~\\
    {\rm level-1:}~~~l \oplus \tau_{\a\b} \oplus \x_{\un{a}}~~,~~\\
    {\rm level-2:}~~~\psi_{\a}\oplus \epsilon_{\a} \oplus L^{(1)}{}_{\a}{}_{\un b}~~,~~ \\
    {\rm level-3:}~~~ \sigma \oplus \omega_{\a\b} \oplus L^{(2)}{}_{\un{a}}~~,\\
    {\rm level-4:}~~~\eta_{\a}~~,
    \end{cases}
\end{equation}
which is consistent with Eq. (\ref{equ:4Dcomp_spinorSF}), upon considering the following fields (which are set to zero
in the gauge $ \Bar{\rD}_{\Dot{\a}} L_{\a} = 0$)
\begin{equation}
\begin{gathered}
    l_{\a} ~=~ L_{\a} | ~~~, \\
    \t_{\a\b} ~=~ \fracm{1}{2} \, \rD_{(\a} L_{\b)} | ~~~,~~~ 
    l ~=~ {\rD}^{\a} L_{\a} | ~~~, \\
   \psi_{\a} ~=~ \rD^{2} L_{\a} | ~~~.
\end{gathered}
\label{equ:4Dcomp_spinorSFb}
\end{equation}

Similarly, for the superfield $ \Bar{L}_{\Dot\a}$, the Dynkin Label and YT descriptions of its component fields are as below.
Its component content is actually ${\cal V}\otimes\CMTred{[1,0]}$, since the subscript $\Dot\a$ is equivalent to the conjugate of the superscript $\Dot\a$, i.e. superscript $\a$. 

\begin{equation}
    \Bar{L}_{\Dot\a} ~=~
    \begin{cases}
    {\rm level-0:}~~~\CMTred{[1,0]}~~,~~\\
    {\rm level-1:}~~~\CMTB{[0,0]} \oplus \CMTB{[2,0]} \oplus\CMTB{[1,1]} ~~,~~\\
    {\rm level-2:}~~~\CMTred{[0,1]}\oplus(2)\CMTred{[1,0]}\oplus\CMTred{[2,1]}~~,~~ \\
    {\rm level-3:}~~~ \CMTB{[0,0]} \oplus \CMTB{[2,0]} \oplus \CMTB{[1,1]}~~,\\
    {\rm level-4:}~~~\CMTred{[1,0]}~~.
    \end{cases}
\end{equation}

\begin{equation}
    \Bar{L}_{\Dot\a} ~=~
    \begin{cases}
    {\rm level-0:}~~~\CMTred{\ytableaushort{\tinytwo}}~~,~~\\
    {\rm level-1:}~~~\CMTB{\singlet} \oplus {\CMTB{\ydiagram{1,1}}}_{\rm IR,-} \oplus {\CMTB{\ydiagram{1}}}~~,~~\\
    {\rm level-2:}~~~(2)\CMTred{\ytableaushort{\tinytwo}}\oplus\CMTB{\ydiagram{1}}\otimes\CMTred{\ytableaushort{\tinytwo}}~~,~~ \\
    {\rm level-3:}~~~ \CMTB{\singlet} \oplus {\CMTB{\ydiagram{1,1}}}_{\rm IR,-} \oplus {\CMTB{\ydiagram{1}}}~~,\\
    {\rm level-4:}~~~\CMTred{\ytableaushort{\tinytwo}}~~.
    \end{cases}
    \label{eqn:LalphadotCompYT}
\end{equation}

Finally, the index notation of its component fields is presented below.

\begin{equation}
    \Bar{L}_{\Dot\a} ~=~
    \begin{cases}
    {\rm level-0:}~~~\Bar{l}_{\Dot\a}~~,~~\\
    {\rm level-1:}~~~\Bar{l} \oplus \Bar{\tau}_{\Dot\a\Dot\b} \oplus \Bar{\x}_{\un{a}}~~,~~\\
    {\rm level-2:}~~~\Bar{\psi}_{\Dot\a}\oplus \Bar{\epsilon}_{\Dot\a} \oplus \Bar{L}^{(1)}{}_{\Dot\a}{}_{\un b}~~,~~ \\
    {\rm level-3:}~~~ \Bar{\sigma} \oplus \Bar{\omega}_{\Dot\a\Dot\b} \oplus \Bar{L}^{(2)}{}_{\un{a}}~~,\\
    {\rm level-4:}~~~\Bar{\eta}_{\Dot\a}~~.
    \end{cases}
\end{equation}

\section{From Superfields to Adynkrafields}
\label{sec:AGOvrVw}

\subsection{Adynkra Genomes}

$~~~~$ As mentioned above, adynkras correspond to a ``2.0 version'' of adinkras.  Thus, it 
is natural for adinkras and adynkras to share some properties.  Many years ago in the study 
of the (dimensionless) adinkra adjacency matrices ${\bm {\rL}}{}_{{}_{\rI}}$ and ${\bm {\rR}
}{}_{{}_{\rI}}$ \cite{ULTR}, it was shown that adinkra nodes could be raised/lowered 
to connect distinct supermultiplets. This gives rise to the concept of a ``level parameter" 
whose change connects different supermultiplets.  

The physical reason for the introduction of the  ``level parameter" is to account for the diverse 
engineering dimensions of component fields in the usual $\theta$-expansion present in 
superfields. The level parameter also accounts for the compatibility of component fields with 
distinct engineering dimensions to occur in the same adinkra.  In the context of adynkras, 
multiple level parameters allow for the same YT to occur multiple times within a single level.

Compared with the conventional superfield approach, the adynkra approach offers two 
advantages: coding and analytical computation: 
\vskip0.01in
\noindent
(a.) as shown in the works of \cite{Cntg-11D,Cntg-10D,Cntg-method,Cntg-lowD}, codes 
have be written and practically 
\newline \noindent $~~~~~~$ 
executed on existing IT platforms to derive component-level results, and
\newline \noindent
(b.) these methods can determine the non-vanishing terms in Fierz identities, 
\newline \noindent $~~~~~~$ 
although not necessarily their normalization constants.

\noindent
Both of these advantages are derived from the same source.  Restricting the domain of 
exploration in terms of either YT or equivalently DLs, permits the harnessing of powerful 
mathematical results of long standing.

At this juncture, several examples are in order to show the calculational efficacy and efficiency 
of the Adynkra Genome concept.

We shall first review the prescription given in \cite{Cntg-method} for how to construct an 
adynkrafield.  The first operator required is an ``Adynkra Genome'' (AG).  We will use the 
symbol $\AG$ to denote such an AG that will act on the space of YT.  There are several 
parts to its construction.

We introduce ``level parameters'' $\ell$ and $\Bar{\ell}$ where the former is associated with the
$\CMTred{[1,0]}$ irrep and the latter with the $\CMTred{[0,1]}$ irrep. Recalling the results of 
Eq.\ (\ref{equ:SYTbasic_4D}), in a respective graphical manner, the DLs are associated 
with $\CMTred{\ytableaushort{\tinytwo}}$ and $\CMTred{\ytableaushort{\tinytwobar}}$.

To see the mathematical meaning of the AG\,\footnote{For a more detailed discussion in 
higher-dimensional superspaces, we refer the reader to Section 8.4 of \cite{Cntg-method}.}, 
we consider a quantity $\bm{\cal K}$ from a superfield. It can be obtained by ``stripping" 
a scalar superfield of all of its field components. Namely,  
\begin{equation}
    \bm{\cal K} ~=~ 1 ~\oplus~\theta^{\a} ~\oplus~ \bar{\theta}^{\dot\a}~\oplus~\theta^2 ~\oplus~\bar{\theta}^2 ~\oplus~\theta^{\a}\bar{\theta}^{\dot\a} ~\oplus~ \cdots ~~.
\end{equation}
Up to some normalization factors, the $\bm{\cal K}$ and Adynkra Genome $\AG$ are related to one another via a change of basis $\theta^{\a}\to\ell \,  \CMTred{\ytableaushort{\tinytwo}}$ 
and $\bar{\theta}^{\dot\a}\to\Bar{\ell} \,  \CMTred{\ytableaushort{\tinytwobar}}$, where the engineering dimension is encoded in the level parameter and the algebraic structure is presented by YTs. 

The next step is to form the products $\ell \,  \CMTred{\ytableaushort{\tinytwo}}$ 
and $\Bar{\ell} \,  \CMTred{\ytableaushort{\tinytwobar}}$ and exponentiate them 
to obtain
\begin{equation}
\begin{split}
    \AG^{(L)} \big[\, \ell \, \CMTred{\ytableaushort{\tinytwo}} \,\big] ~=&~ \exp \big[\, \ell \, \CMTred{\ytableaushort{\tinytwo}} \,\big]  ~~~, \\
    \AG^{(R)} \big[\, \Bar{\ell} \, \CMTred{\ytableaushort{\tinytwobar}} \,\big] ~=&~ \exp \big[\, \Bar{\ell} \, \CMTred{\ytableaushort{\tinytwobar}} \,\big]  ~~~,
\end{split}
\label{eq:G-Exp}
\end{equation}
which permit us to define the AG's via the definitions
\begin{equation}
    \AG \big[\, \ell , \Bar{\ell} , \CMTred{\ytableaushort{\tinytwo}} , \CMTred{\ytableaushort{\tinytwobar}} \,\big] ~ [ \CMTB{YT} ]
    ~\equiv~
    \AG^{(L)} \big[\, \ell \, \CMTred{\ytableaushort{\tinytwo}} \,\big] ~ \AG^{(R)} \big[\, \Bar{\ell} \, \CMTred{\ytableaushort{\tinytwobar}} \,\big] ~ [ \CMTB{YT} ] ~~~,
\label{eq:G-BLU}
\end{equation}
for a bosonic YT $\left[ \CMTB{YT} \right]$ and
\begin{equation}
    \AG \big[\, \ell , \Bar{\ell} , \CMTred{\ytableaushort{\tinytwo}} , \CMTred{\ytableaushort{\tinytwobar}} \,\big] ~ [ \CMTred{YT} ]
    ~\equiv~
    \AG^{(L)} \big[\, \ell \, \CMTred{\ytableaushort{\tinytwo}} \,\big] ~ \AG^{(R)} \big[\, \Bar{\ell} \, \CMTred{\ytableaushort{\tinytwobar}} \,\big] ~ [ \CMTred{YT} ] ~~~,
\label{eq:G-RED}
\end{equation}
for a fermionic YT $\left[ \CMTR{YT} \right]$.
\vskip0.10in 
\noindent
There are a few points about these equations to note:
\vskip0.4pt 
\noindent
(a.) the rules for construction of $\left[ \CMTB{YT} \right]$ and 
$\left[ \CMTR{YT} \right]$ have been stated in cases for higher
\newline \noindent $~~~~~~$ dimensional theories in the works of \cite{Cntg-11D,Cntg-10D,Cntg-method,Cntg-lowD,Cntg-Weyl} and may be
adapted to the \newline \noindent $~~~~~~$
present case in the most obvious ways, and
\newline \noindent
(b.) 
the rules of multiplications for ${\CMTred{\ytableaushort{\tinytwo}}}$ and 
${\CMTred{\ytableaushort{\tinytwobar}}}$ needed for the exponential expansions
\newline \noindent $~~~~~~$  
of the fermionic YT are {\em {not}} the usual ones that correspond to  
\be
\begin{split}
{\CMTR{\ytableaushort{\tinytwo}}} ~\otimes~ {\CMTR{\ytableaushort{\tinytwo}}} ~=&~ \CMTR{\ytableaushort{\tinytwo,\tinytwo}} ~  
\oplus ~  \CMTR{\ytableaushort{\tinytwo\tinytwo}}  ~=~ \CMTB{\singlet} ~  
\oplus ~ {\CMTB{\ydiagram{1,1}}}_{\rm IR,-} ~~~,\\
{\CMTR{\ytableaushort{\tinytwobar}}} ~\otimes~ {\CMTR{\ytableaushort{\tinytwobar}}} ~=&~ \CMTR{\ytableaushort{\tinytwobar,\tinytwobar}} ~  
\oplus ~  \CMTR{\ytableaushort{\tinytwobar\tinytwobar}}  ~=~ \CMTB{\singlet} ~  
\oplus ~ {\CMTB{\ydiagram{1,1}}}_{\rm IR,+} ~~~,
\label{eq:notWdG}
\end{split}
\ee
\noindent $~~~~~~$ 
instead we have
\be
{\CMTR{\ydiagram{1}}} ~ \wedge ~{\CMTR{\ydiagram{1}}} ~=~ {\CMTR{\ydiagram{1,1}}} ~=~ \CMTB{\singlet}
~~~,
\label{eq:WdG}
\ee
\newline \noindent $~~~~\,~~~$  
which is valid whether both the red tableaux in 
Eq.\ (\ref{eq:WdG}) 
\newline \noindent $~~~~\,~~~$
correspond to $\CMTred{\ytableaushort{\tinytwo}}$ or to
$\CMTred{\ytableaushort{\tinytwobar}}$ .
\vskip10pt \indent
Finally, we believe it is worth pausing here to make a point about the adynkra technology we have
described in Eq.\ (\ref{eq:notWdG}).  In a sense, these two equations are examples of the `secret 
identities' of adynkras that drive their computational efficiencies, in comparison to Salam-Strathdee
superfields, for identifying the Lorentz representations of the component fields contained within them.  
This is a true statement not just in these explicit examples, but in all calculations involving adynkras.

The equivalent statements for superfield expansions are the Fierz identities.  These identities do two
things:
(a.) they identify the bosonic representations that emerge from products of spinor representations, and
(b.) they contain information about Clebsch-Gordan coefficients.

The results in Eq.\ (\ref{eq:notWdG}) only contain information about the bosonic representations, but contain
no information about Clebsch-Gordan coefficients.  Thus, adynkrafields do not replace Salam-Strathdee 
superfields.  Instead, each has complementary roles.

There is one other important feature to observe in these constructions.  The presence of the ``level parameters'' 
(i.e. $\ell$ and ${\Tilde {\ell}}$) is required from two distinct viewpoints. 

\subsection{Adynkra Genome Examples}
\label{sec:AGEb}

$~~~~$ In this section, we will use the same six supermultiplets to be discussed in section \ref{sec:AE} 
and explicitly demonstrate the corresponding adynkra genomes in the same order of presentation.

\begin{align}
    \AG \big[\, \ell , 0 , \CMTred{\ytableaushort{\tinytwo}} , \CMTred{\ytableaushort{\tinytwobar}} \,\big] ~ \CMTB{\singlet} 
    ~=&~ \AG^{(L)} \big[\, \ell \, \CMTred{\ytableaushort{\tinytwo}} \,\big] ~ \CMTB{\singlet} 
    ~=~ \CMTB{\singlet} ~\oplus~ \ell \, \CMTred{\ytableaushort{\tinytwo}} ~\oplus~ \fracm{1}{2!} \, (\ell)^{2} ~ \CMTB{\singlet} ~~~,
    \label{equ:GnCS} \\[1em]
    \AG \big[\, \ell , 0 , \CMTred{\ytableaushort{\tinytwo}} , \CMTred{\ytableaushort{\tinytwobar}} \,\big] ~ \CMTred{\ytableaushort{\tinytwo}}
    ~=&~ \AG^{(L)} \big[\, \ell \, \CMTred{\ytableaushort{\tinytwo}} \,\big] ~ \CMTred{\ytableaushort{\tinytwo}} 
    ~=~ \CMTred{\ytableaushort{\tinytwo}} ~\oplus~ \ell ~ \big(\, \CMTB{\singlet} \,\oplus\, {\CMTB{\ydiagram{1,1}}}_{\rm IR,-} ~\big) ~\oplus~ \fracm{1}{2!} \, (\ell)^{2} ~ \CMTred{\ytableaushort{\tinytwo}} ~~~,
    \label{equ:GnCSp1} \\[1em]
    \AG \big[\, \ell , 0 , \CMTred{\ytableaushort{\tinytwo}} , \CMTred{\ytableaushort{\tinytwobar}} \,\big] ~ \CMTred{\ytableaushort{\tinytwobar}}
    ~=&~ \AG^{(L)} \big[\, \ell \, \CMTred{\ytableaushort{\tinytwo}} \,\big] ~ \CMTred{\ytableaushort{\tinytwobar}} 
    ~=~ \CMTred{\ytableaushort{\tinytwobar}} ~\oplus~ \ell \, {\CMTB{\ydiagram{1}}} ~\oplus~ \fracm{1}{2!} \, (\ell)^{2} ~ \CMTred{\ytableaushort{\tinytwobar}} ~~~,
    \label{equ:GnCSp2}
\end{align}
and 
\begin{align}
\begin{split}
    \AG \big[\, \ell , \Bar{\ell} , \CMTred{\ytableaushort{\tinytwo}} , \CMTred{\ytableaushort{\tinytwobar}} \,\big] ~ \CMTB{\singlet} 
    ~=~ \CMTB{\singlet} 
    &~\oplus~ \ell ~ \CMTred{\ytableaushort{\tinytwo}}  
    ~\oplus~ \Bar{\ell} ~ \CMTred{\ytableaushort{\tinytwobar}}   ~\oplus~ \fracm{1}{2!}\,(\ell)^2 \, \CMTB{\singlet} 
    ~\oplus~ \fracm{1}{2!}\,(\Bar{\ell})^2 \, \CMTB{\singlet}  
    ~\oplus~ \ell\,\Bar{\ell} ~ {\CMTB{\ydiagram{1}}} \\
    &~\oplus~ \fracm{1}{2!}\,(\ell)^2 \, \Bar{\ell} ~ \CMTred{\ytableaushort{\tinytwobar}} 
    ~\oplus~ \fracm{1}{2!}\,(\Bar{\ell})^2\,\ell ~ \CMTred{\ytableaushort{\tinytwo}} 
    ~\oplus~ \fracm{1}{2!2!}\,(\ell)^2\,(\Bar{\ell})^2 \, \CMTB{\singlet}  ~~~, 
\end{split}
\label{equ:GnVS} \\[1em]
\begin{split}
    \AG \big[\, \ell , \Bar{\ell} , \CMTred{\ytableaushort{\tinytwo}} , \CMTred{\ytableaushort{\tinytwobar}} \,\big] ~ \CMTred{\ytableaushort{\tinytwo}} 
    ~=~ \CMTred{\ytableaushort{\tinytwo}}
    &~\oplus~ \ell ~ \big(\, \CMTB{\singlet} \,\oplus\, {\CMTB{\ydiagram{1,1}}}_{\rm IR,-} ~\big)  
    ~\oplus~ \Bar{\ell}\, \,{\CMTB{\ydiagram{1}}}
    ~\oplus~ \fracm{1}{2!}\,(\ell)^2\, \CMTred{\ytableaushort{\tinytwo}}
    ~\oplus~ \fracm{1}{2!}\,(\Bar{\ell})^2\, \CMTred{\ytableaushort{\tinytwo}}  \\
    &~\oplus~ \ell\,\Bar{\ell}\,  \big(~ \CMTred{\ytableaushort{\tinytwobar}} \,\oplus\, {\CMTB{\ydiagram{1}}\CMTred{\ytableaushort{\tinytwo}}}_{\rm IR} \,\big) 
    ~\oplus~ \fracm{1}{2!}\,(\ell)^2\, \Bar{\ell}\,  \,{\CMTB{\ydiagram{1}}}
    ~\oplus~ \fracm{1}{2!}\,(\Bar{\ell})^2\,\ell\, \big(\, \CMTB{\singlet} \,\oplus\, {\CMTB{\ydiagram{1,1}}}_{\rm IR,-} ~\big) \\
    &~\oplus~ \fracm{1}{2!2!}\,(\ell)^2\,(\Bar{\ell})^2\, \CMTred{\ytableaushort{\tinytwo}} ~~~,
\end{split}
\label{equ:GnVSSp} \\[1em]
\begin{split}
    \AG \big[\, \ell , \Bar{\ell} , \CMTred{\ytableaushort{\tinytwo}} , \CMTred{\ytableaushort{\tinytwobar}} \,\big] ~ \CMTB{\ydiagram{1}}
    ~=~ {\CMTB{\ydiagram{1}}}
    &~ \oplus ~ \ell\,  \big( \,\CMTred{\ytableaushort{\tinytwobar}} \,\oplus\, {\CMTB{\ydiagram{1}}\CMTred{\ytableaushort{\tinytwo}}}_{\rm IR}  \, \big) 
    ~ \oplus ~ \Bar{\ell}\,  \big( \,\CMTred{\ytableaushort{\tinytwo}} \,\oplus\, {\CMTB{\ydiagram{1}}\CMTred{\ytableaushort{\tinytwobar}}}_{\rm IR}  \, \big) 
    ~ \oplus ~ \fracm{1}{2!}\,(\ell)^2\, \,{\CMTB{\ydiagram{1}}} \\
    &~ \oplus ~ \fracm{1}{2!}\,(\Bar{\ell})^2\, \,{\CMTB{\ydiagram{1}}} 
    ~ \oplus ~ \ell\,\Bar{\ell}\, \big(\, {\CMTB{\ydiagram{2}}}_{\rm {IR}} \, \oplus{\CMTB{\ydiagram{1,1}}}_{\rm IR,-}\, \oplus \, {\CMTB{\ydiagram{1,1}}}_{\rm IR,+} \, \oplus\, \CMTB{\singlet} \,\big) \\
    &~ \oplus ~ \fracm{1}{2!}\,(\ell)^2\, \Bar{\ell}\, 
    \big( \,\CMTred{\ytableaushort{\tinytwo}} \,\oplus\, {\CMTB{\ydiagram{1}}\CMTred{\ytableaushort{\tinytwobar}}}_{\rm IR}  \, \big) 
    ~ \oplus ~ \fracm{1}{2!}\,(\Bar{\ell})^2\,\ell\,  \big( \,\CMTred{\ytableaushort{\tinytwobar}} \,\oplus\, {\CMTB{\ydiagram{1}}\CMTred{\ytableaushort{\tinytwo}}}_{\rm IR}  \, \big) \\
    &~ \oplus ~ \fracm{1}{2!2!}\,(\ell)^2\,(\Bar{\ell})^2\, \,{\CMTB{\ydiagram{1}}} ~~~.
\end{split}
\label{equ:GnVV}
\end{align}

Each of these adynkra genomes will shortly be shown to correspond to a previously identified 4D,
$\cal N$ = 1 supermultiplet in the literature with the following correspondences:
\vskip0.010in \indent 
(1.) chiral supermultiplet - (\ref{equ:GnCS}),
\newline \indent 
(2.) 2-form gauge field supermultiplet
- (\ref{equ:GnCSp1}),
\newline \indent 
(3.) 1-form variant gauge field supermultiplet
- (\ref{equ:GnCSp2}),
\newline \indent 
(4.) 1-form gauge field supermultiplet - (\ref{equ:GnVS}),
\newline \indent 
(5.) matter gravitino - (\ref{equ:GnVSSp}), and
\newline \indent 
(6.) supergravity supermultiplet - (\ref{equ:GnVV}).
\vskip0.010in \noindent
In the cases of (\ref{equ:GnVS}) and
(\ref{equ:GnVV}), we can respectively look back at the equations and compare them with (\ref{eqn:VCompDyn}) and (\ref{eqn:HaCompDyn})
and recognize that the adynkra genomes are expansions in the level parameter where the latter take the place of the usual $\theta$-expansions.
In both cases, exactly identical sets of representations of the Lorentz groups show up at corresponding orders in the respective expansions.

\section{\texorpdfstring{Adynkra Genomes to
4D, ${\cal N}=1$ Adynkrafields}{Adynkra Genomes to
4D, N=1 Adynkrafields}}
\label{sec:AGbEbb}

In this section, we transition from adynkra genomes to the introduction of adynkrafields. Before proceeding, we first summarize the key concepts of the adynkra formalism.\,\footnote{Although these ideas and definitions have been reviewed in previous sections, we refer the reader to \cite{Cntg-11D,Cntg-10D,Cntg-method,Cntg-lowD,Cntg-Weyl} for more detailed discussions and examples in various spacetime dimensions.}
Table~\ref{tab:summary} lists these essential concepts and contrasts them with their analogs in the superspace formalism.
\begin{table}[t]
    \centering
    \begin{tabular}{c|c}
        Superspace formalism & Adynkra formalism \\\hline\hline
        Grassmann variable $\theta$ & $\ell\,\CMTred{\ydiagram{1}}$\,, where the level parameter $\ell$ carries \\
        &   the engineering dimension and the spinorial YT $\CMTred{\ydiagram{1}}$ \\
        & denotes the algebraic structure
        \\\hline
        Component fields & field variables $\Phi_{\rm [idx]}(x)$, which correspond to \\ 
        & irreducible representations of the Lorentz group. \\\hline
        & The index structures [idx],  Dynkin Labels, and \\
        & bosonic/spinorial YTs are in one-to-one correspondence.  \\\hline
        Irreducible $\theta$-monomials & level parameters + irreducible YTs \\\hline
        $\bm{\cal K} = 1 \oplus_{p=1}^d\theta^{\a_1}\cdots\theta^{\a_p}$ & Adynkra genome $\AG$ \\\hline
        Superfield decomposition of $\Phi_{\rm [idx]}(x,\theta)$ & Adynkrafield $\widehat{\Phi}_{\rm [idx]}(x)$ \\\hline
    \end{tabular}
    \caption{Summary of key concepts in the adynkra formalism (for a generic spacetime dimension)}
    \label{tab:summary}
\end{table}

\subsection{Representation Theory Description Of Adynkra Genomes}
\label{sec:AGC}

$~~~~$ In formal mathematical language, adinkras are ``forget functors" of adynkras.  As we
will see nodes in adynkras graphically carry spin information about the higher dimensional 
fields they describe while no such information is carried by the nodes of adinkras.  This is 
achieved by augmenting the nodes in adinkras with additional information in the form 
of Dynkin Labels or Young Tableaux.

The main advantage of this will arise because this means the sets of nodes of adynkras,
while not by themselves forming mathematical rings -- as the latter must realize the two 
operations of addition and multiplication,  can serve as a basis or labelling set for rings
that arise in representation theory.

The basis of adinkras is formed by hypercubes in arbitrary dimensions.  In fact, these
hypercubes are Hamming cubes where vertices are all binary strings of a fixed length
$n$ and thus representable as
\begin{equation}
\{ 0. \, 1 \}^n
\label{equ:RepTh1}
\end{equation}
with two vertices connected by an edge when their strings differ by exactly one position.
Thus, these $n$-dimensional Boolean hypercubes may be regarded as the generators
of adinkras.  However, true adinkras only arise when a ``height function,'' of the nature
that is seen in discussions of Riemann surfaces is included.  Thus, these $n$-dimensional 
Boolean hypercubes are forgetful functors of adinkras.  These mathematical objects
can be thought of as the ``generators'' of adinkras.

Thus, a different conceptual foundation is needed for the Adynkra concept to replace
the $n$-dimensional Boolean hypercubes seen at the basis of adinkras.  This is 
accomplished by the introduction of the ``Adynkra Genome Concept.''  We generically
denote such an object by the symbol $\AG $.  These objects should be related to
adynkras as the $n$-dimensional Boolean hypercubes are related to adinkras.

For a generic superfield in a generic superspace\footnote{Here we have in mind
a superspace where the number of spinor components is d and the number of bosonic
components is of the manifold is D.}, this is accomplished through the definition
\begin{equation}
\AG [\, {d} , \, {\rm D} , \, \ell, \, {\ydiagram{1}} \,] ~\define ~ {\cal P}_{\rm D} \,\,
{\exp}{}_{{\wedge}^{d-1}} {\left[ \, \ell \,   {\ydiagram{1}} \, \right] } ~~~,
\label{equ:RepTh2}
\end{equation}
where the arguments of the genome on the left hand side are:\\
\noindent\begin{minipage}[t]{15cm}
\begin{itemize}
    \item[] $d$ - the number of components of the fundamental fermionic representation \newline \indent \hskip0.25in associated with the manifold,
    \item[] D - the number of components of the bosonic coordinates \newline \indent \hskip0.28in associated with the manifold,
    \item[] $\ell$ - the level parameter as seen in the example of the 4D theory, and
    \item[] ${\ydiagram{1}} $ - a Young Tableau associate with the $A_{d-1}$ series of Cartan's Lie \newline \indent \hskip0.33in algebra classification's fundamental representation. 
\end{itemize}
\end{minipage}\\

\noindent
As the Young Tableau on the left hand side is distinct from the ones seen previously, its 
coloring is black to prevent confusion with either the blue or red ones previously introduced.

On the right hand side of the equation, the quantity ${\cal P}_{\rm D} $ is projection operator 
that `projects' all black YT's ${\ydiagram{1}}$ onto either blue YT's $\CMTB{\ydiagram{1}}$ or red 
YT's $\CMTR{\ydiagram{1}}$ depending on whether the coefficient that proceeds the factor
containing the tableau is and even or odd power of $\ell$.  Computationally, this is where the
using of branching rules is implemented.  These replace the use of Fierz identities to uncover
the explicit spin of component fields.

The expression ${\exp}{}_{{\wedge}^{d-1}}$ denotes the usual expansion of the exponential function,
however where the `wedge' product of the all black YT's (${\ydiagram{1}}$) are to be used as the
multiplication rule.  It is to be taken up to a maximum of $d-1$ times between black YT's. Thus, the 
explicit expression as 
\begin{equation} 
\begin{split}
{\exp}{}_{{\wedge}^{d-1}} {\left[ \, \ell \,   {\ydiagram{1}}\, \right] } ~=~ 
& 
{ {\CMTB { \cdot}}} ~\oplus~ \ell \,  {\ydiagram{1}}~\oplus~ \fracm {1}{2!} (\ell)^2 \, \left( \, {\ydiagram{1}}\wedge  {\ydiagram{1}}
 \, \right)    ~\oplus~ \fracm {1}{3!} (\ell)^3 \,  \left( \,{\ydiagram{1}}\wedge  {\ydiagram{1}} \wedge  {\ydiagram{1}}\, \right)  \\
 & 
 ~~\oplus~ \fracm {1}{4!} (\ell)^4 \,  \left( \,{\ydiagram{1}}\wedge  {\ydiagram{1}} \wedge  {\ydiagram{1}}  \wedge  {\ydiagram{1}}\, \right) ~\oplus~ \cdots \\
 & 
 ~~\oplus~ \fracm {1}{d!} (\ell)^d \,  \left( \,{\ydiagram{1}} \wedge~ \cdots \,  (d-1) {\rm {times}}  \, \cdots ~ \wedge  {\ydiagram{1}}\, \right)    ~~~,
 \end{split}
 \label{equ:RepTh3}
\end{equation}
arises.  We believe it is extremely satisfying that representation theory equations (\ref{equ:RepTh2}) and (\ref{equ:RepTh3})
allow for such a concise definition the multiple component fields that arise in the superfield concept of 
Salam and Strathdee.  Now we return to the case were $d$ = D = 4 as appropriate in our prior chapters.

Clearly, the definitions of adynkra genomes in the equations (\ref{eq:G-BLU}) and (\ref{eq:G-RED}) 
imply that these mathematical objects may be regarded as elements in the space of expansions 
over YT or alternatively the space of DLs.  Passing from an adynkra genome to an adynkrafield amounts
to taking an ``overlap'' between an adynkra genome  $\AG \big[\, \ell , \Bar{\ell} , \CMTred{\ytableaushort{\tinytwo}} , \CMTred{\ytableaushort{\tinytwobar}} \,\big]$, specified by $[ {\CMTB {YT}}] $ and  $[ {\CMTR {YT}} ] $,
and the space of field variables denoted by$\{ {\cal F} \} $.  

It is well known that the space of fields $ \{ {\cal F} \}$ describing all representations of the Lorentz 
group is bisected into two parts and can be described by the equation 
\be {\bm { \{ {\cal F} \} } } ~=~ {\bm { \{ {\cal F} \} } }_{b} ~ \oplus~   {\bm { \{ {\cal F} \} } }_{f} ~~~,
\label{eq:FLdsp8c} \ee
where ${\bm { \{ {\cal F} \} } }_{b}$ and ${\bm { \{ {\cal F} \} } }_{f}$
denote, respectively, the spaces of bosonic and
fermionic representations.  For the convenience of the reader, we have appropriately adapted  
Eq.\ (\ref{equ:Fsp8e}) (showing conventional displays of fields) and Eq.\ (\ref{equ:Fsp8cYT}) 
(showing YT-field pairings) 
\begin{equation}
{\bm { \{ {\cal F} \} } }  =  \{ 
\begin{matrix} \text{scalar} \\ \phi(x) \\ \text{spin}-0, \end{matrix} ,  
\begin{matrix} \text{photon} \\ A_{\un{a}}(x) \\ \text{spin}-1, \end{matrix} ,  
\begin{matrix} \text{graviton} \\ h_{\un{a}\un{b}}(x) \\ \text{spin}-2, \dots\end{matrix} ,
\dots \}  \oplus  \{
\begin{matrix} \text{spinor} \\ \l_{\a}(x) \\ \text{spin}-\frac{1}{2}, \end{matrix} ,  
\begin{matrix} \text{spinor} \\ \l_{\Dot \a}(x) \\ \text{spin}-\frac{1}{2}, \end{matrix} ,  
\begin{matrix} \text{gravitino} \\ \psi_{\un{a}}{}^{\b}(x) \\ \text{spin}-\frac{3}{2}, 
\end{matrix} ,  
\begin{matrix} \text{gravitino} \\ \psi_{\un{a}}{}^{\Dot \b}(x) \\ \text{spin}-\frac{3}{2}, 
\dots
\end{matrix} ,
\dots \}    ,
\label{equ:Fsp8e}
\end{equation}
and our mapping to YT
\begin{equation}
{\bm { \{ {\cal F} \} } }  =  \{ 
\begin{matrix} \text{scalar} \\ \CMTB{\singlet} (x) \\ \text{spin}-0, \end{matrix} ,  
\begin{matrix} \text{photon} \\ \CMTB{\ydiagram{1}}(x) \\ \text{spin}-1, \end{matrix} ,  
\begin{matrix} \text{graviton} \\ \CMTB{\ydiagram{2}}(x) \\ \text{spin}-2, \dots \end{matrix} ,
\dots \}  \oplus  \{
\begin{matrix} \text{spinor} \\ \CMTred{\ytableaushort{\tinytwo}}(x) \\ \text{spin}-\frac{1}{2}, \end{matrix} ,  
\begin{matrix} \text{spinor} \\ \CMTred{\ytableaushort{\tinytwobar}}(x) \\ \text{spin}-\frac{1}{2}, \end{matrix} ,  
\begin{matrix} \text{gravitino} \\ \CMTB{\ydiagram{1}}\CMTred{\ytableaushort{\tinytwobar}}(x) \\ \text{spin}-\frac{3}{2}, \end{matrix} ,  
\begin{matrix} \text{gravitino} \\ \CMTB{\ydiagram{1}}\CMTred{\ytableaushort{\tinytwo}}(x) \\ \text{spin}-\frac{3}{2},\dots \end{matrix} ,
\dots \}    ,
\label{equ:Fsp8cYT}
\end{equation}
motivated by the work in \cite{Cntg-method}, to display more explicitly the contents of 
${\bm { \{ {\cal F} \} } }_{b}$ and ${\bm { \{ {\cal F} \} } }_{f}$.

After making a choice of YT (indicated by a subscript $[{\CMTB {YT}}]/[{\CMTR {YT}}]$), it becomes 
possible to define an adynkrafield denoted by $ {\bm {{ \widehat {\cal F}  }} }(x) $ via 
\be
    {\bm {\widehat{\cal F}} } (x) ~=~ {\Big <} ~ \AG \big[\, \ell , \Bar{\ell} , \CMTred{\ytableaushort{\tinytwo}} , \CMTred{\ytableaushort{\tinytwobar}} \,\big] ~,~  {\bm { \{ {\cal F} \} } } \, {\Big  >} {\Big |}_{[{\CMTB {YT}}]/[{\CMTR {YT}}]} ~~~,
\ee 
where the bracket pairs the YTs appearing in each term of the genome with the corresponding \textit{conjugate} field variables in ${\bm { \{ {\cal F} \} } }$. 
Recall that there is a one-to-one correspondence among irreducible representations, YTs, and index structures~\cite{Cntg-method}. Here, a conjugate field variable is defined as one carrying the index structure associated with the conjugate YT. Moreover, by construction, ${\bm {\widehat{\cal F}} } (x)$ carries no explicit Lorentz indices.  

The field variables selected by the adynkrafield ${\bm {\widehat{\cal F}} } (x)$ are precisely the component fields of the corresponding superfield. This superfield carries the index structure associated with the \textit{conjugate} of $[{\CMTB {YT}}]/[{\CMTR {YT}}]$ following the correspondence between Lorentz index structures and irreducible representations reviewed in Section~\ref{sec:SFrv} and Refs.~\cite{Cntg-method,Cntg-lowD}. 
To make the index structure manifest, one may alternatively define an adynkrafield $\widehat{\Phi}_{\rm [idx]}(x)$ carrying explicit Lorentz indices via
\begin{equation}
   [\overline{\rm idx}]\otimes \widehat{\Phi}_{\rm [idx]}(x) ~=~  {\Big <} ~ \AG \big[\, \ell , \Bar{\ell} , \CMTred{\ytableaushort{\tinytwo}} , \CMTred{\ytableaushort{\tinytwobar}} \,\big] ~,~  {\bm { \{ {\cal F} \} } } \, {\Big  >} {\Big |}_{[{\CMTB {YT}}]/[{\CMTR {YT}}]} 
\end{equation}
where $[\overline{\rm idx}]$ denotes the conjugate irreducible representation/YT associated with the index structure ${\rm [idx]}$. 

Due to the expansions seen in the previous chapter and the pairing rule indicated immediately
above, the adynkrafield becomes a specified expansion with coefficients that are elements of
 $ {\bm { \{ {\cal F} \} } } $.

In the following, the presentation turns to examples.

\subsection{\texorpdfstring{4D, ${\cal N}=1$ Adynkrafield Examples}{4D, N=1 Adynkrafield Examples}}
\label{sec:AE}

$~~~~$ Let us provide a notational note before we dive into adynkrafields.
One notices that in Chapter \ref{sec:AGOvrVw}, when we discuss adynkra genomes, we use $\oplus$ throughout. However, when we translate to adynkrafield notation, there are contractions between ``hidden'' indices carried by YT and explicit indices carried by field variables. In the following, $+$ signs are used when these contractions occur. 

Eq. (\ref{equ:GnCS}) yields 
\begin{equation}
{\Big <} ~ \AG \big[\, \ell , 0 , \CMTred{\ytableaushort{\tinytwo}} , \CMTred{\ytableaushort{\tinytwobar}} \,\big] ~,~  {\bm { \{ {\cal F} \} } } \, {\Big  >} {\Big |}_{\CMTB{\singlet}} ~=~ \Phi(x) ~+~ \ell\,\CMTred{\ytableaushort{\tinytwo}}\,\Psi_{\a}(x) ~+~ \fracm{1}{2!}\,(\ell)^2 \Phi^{(2)}(x)~~~.
\end{equation}
Alternatively, we have
\begin{equation}
\Hat{\Phi}(x)~=~ \Phi(x) ~+~ \ell\,\CMTred{\ytableaushort{\tinytwo}}\,\Psi_{\a}(x) ~+~ \fracm{1}{2!}\,(\ell)^2  \Phi^{(2)}(x)~~~,
\end{equation}
which corresponds to a chiral scalar superfield. This is the simplest example since tensoring with the singlet representation is trivial. However, once a nontrivial representation is involved (i.e., when the corresponding superfield carries bosonic or fermionic indices), the contractions must be treated with care. 
We will see this in the examples that follow. 

For the genome in Eq. (\ref{equ:GnCSp1}), we first introduce the notation 
\begin{equation}
    \Hat{\AG}\,\left[\,\ell, 0,\CMTred{\ytableaushort{\tinytwo}},\CMTred{\ytableaushort{\tinytwobar}}\,\right]\,\CMTred{\ytableaushort{\tinytwo}} ~=~ {\Big <} ~ \AG \big[\, \ell , 0 , \CMTred{\ytableaushort{\tinytwo}} , \CMTred{\ytableaushort{\tinytwobar}} \,\big] ~,~  {\bm { \{ {\cal F} \} } } \, {\Big  >} {\Big |}_{\CMTB{\CMTred{\ytableaushort{\tinytwo}}}}
\end{equation}
which appends the projected list of field variables to the original genome as
\begin{equation}
\Hat{\AG}\,\left[\,\ell, 0,\CMTred{\ytableaushort{\tinytwo}},\CMTred{\ytableaushort{\tinytwobar}}\,\right]\,\CMTred{\ytableaushort{\tinytwo}}~=~ \CMTred{\ytableaushort{\tinytwo}}\,\Phi_{\a}(x) ~+~ \ell\,\Phi(x) ~+~ \ell\,{\CMTB{\ydiagram{1,1}}}_{\rm IR-}\,\Phi_{{\un a}{\un b}}^{(+)}(x)~+~ \fracm{1}{2!}\,(\ell)^2\,\CMTred{\ytableaushort{\tinytwo}}\,\varphi_{\a}(x)~~~.
\end{equation}
The appended field variables carry index structures corresponding to the conjugates of the associated YTs. For example, $\CMTred{\ytableaushort{\tinytwo}}$ corresponds to an upper-level spinor index $\a$, and therefore the appended field $\Phi_{\a}(x)$ carries a lower-level spinor index $\a$. 
Next, the adynkrafield with the same index structure as the corresponding superfield $\Phi_{\a}(x,\theta)$ can be obtained by removing the YT appearing in the level-0 term (in this example $\CMTred{\ytableaushort{\tinytwo}}$). Equivalently, this amounts to replacing the YTs attached to the level parameters in $\Hat{\AG}$ by those appearing in Eq.~(\ref{equ:GnVS}). Namely, 
\begin{equation}
\Hat{\Phi}_{\a}(x)~=~ \Phi_{\a}(x) ~+~ \ell\,\CMTred{\ytableaushort{\tinytwo}}\,\Phi(x) ~+~ \ell\,\CMTred{\ytableaushort{\tinytwo}}\,\Phi_{\a\b}(x) ~+~ \fracm{1}{2!}\,(\ell)^2\varphi_{\a}(x)~~~.
\label{equ:adynkFchiral}
\end{equation}
In this way, the adynkrafield $\Hat{\Phi}_{\a}(x)$ can indeed be interpreted as a superfield decomposition in which the irreducible $\theta$-monomials are replaced by level parameters and YTs.\,\footnote{The Lorentz indices may appear to be inconsistent in Eq.~(\ref{equ:adynkFchiral}); in particular, the second term on the RHS seems to carry an upper spinor index. This apparent inconsistency arises because the spinor metric $C_{\a\b}$ is omitted in the adynkrafield. Restoring the spinor metric gives
\begin{equation}
    \Hat{\Phi}_{\a}(x)~=~ \Phi_{\a}(x) ~+~ \ell\,\CMTred{\ytableaushort{\tinytwo}}C_{\a\b}\,\Phi(x) ~+~ \ell\,\CMTred{\ytableaushort{\tinytwo}}\,\Phi_{\a\b}(x) ~+~ \fracm{1}{2!}\,(\ell)^2\varphi_{\a}(x) ~~.
\end{equation}
Comparing this expression with the standard chiral superfield decomposition
\begin{equation}
    \Phi_{\a}(x,\theta)~=~ \lambda_{\a}(x) ~+~ \theta^{\b} C_{\a\b} D(x) ~+~ \theta^{\b}f_{\a\b}(x) ~-~ \theta^2\chi_{\a}(x) ~~, 
\end{equation}
one sees that the adynkrafield replaces the $\theta$-monomials by Young Tableaux while encoding the engineering dimensions associated with $\theta$ through the level parameter $\ell$, with the numerical coefficients omitted.
} 

For the genome in Eq. (\ref{equ:GnCSp2}), the corresponding $\Hat{\AG}$ reads
\begin{equation}
\Hat{\AG}\,\left[\,\ell, 0,\CMTred{\ytableaushort{\tinytwo}},\CMTred{\ytableaushort{\tinytwobar}}\,\right]\,\CMTred{\ytableaushort{\tinytwobar}}~=~ \CMTred{\ytableaushort{\tinytwobar}}\,{\Phi}_{\Dot \a}(x) ~+~ \ell\,\CMTB{\ydiagram{1}}\,\Phi_{\un a}(x) ~+~ \fracm{1}{2!}\,(\ell)^2\,\CMTred{\ytableaushort{\tinytwobar}}\,\rho_{\Dot\a}(x)~~~,
\end{equation}
which associates to the superfield ${\Phi}_{\Dot\a}(x,\theta)$.
The corresponding adynkrafield is 
\begin{equation}
\Hat{\Phi}_{\Dot\a}(x)~=~ \Phi_{\Dot\a}(x) ~+~ \ell\,\CMTred{\ytableaushort{\tinytwo}}\,\Phi_{\un a}(x) ~+~ \fracm{1}{2!}\,(\ell)^2\rho_{\Dot\a}(x)~~~.
\end{equation}

For the examples in Eqs.~(\ref{equ:GnVS})--(\ref{equ:GnVV}), we omit the $\Hat{\AG}$ construction and present the final adynkrafields $\hat{\Phi}_{\rm [idx]}(x)$ directly.
Following the same procedure, Eq. (\ref{equ:GnVS}) yields an adynkrafield
\begin{equation}
    \begin{split}
        \Hat{\cal V}(x) ~=&~ V(x) 
        ~+~ \ell\,\CMTred{\ytableaushort{\tinytwo}}\,\Psi_{\a} (x)  
        ~+~ \Bar{\ell}\,\CMTred{\ytableaushort{\tinytwobar}}\,\Psi_{\Dot\a}(x)   
        ~+~ \fracm{1}{2!}\,(\ell)^2\, \Phi(x)  
        ~+~ \fracm{1}{2!}\,(\Bar{\ell})^2\, \Tilde{\Phi}(x)  \\
        &~+~ \ell\,\Bar{\ell}\,{\CMTB{\ydiagram{1}}}\,\Phi_{\un a}(x) 
        ~+~ \fracm{1}{2!}\,(\ell)^2\, \Bar{\ell}\, \CMTred{\ytableaushort{\tinytwobar}}\,\Hat{\Psi}_{\Dot\a}(x) 
        ~+~ \fracm{1}{2!}\,(\Bar{\ell})^2\,\ell\,\CMTred{\ytableaushort{\tinytwo}}\,\Hat{\Psi}_{\a} (x) \\
        &~+~ \fracm{1}{2!2!}\,(\ell)^2\,(\Bar{\ell})^2\, \Hat{V}(x) ~~~.
    \end{split}
    \label{equ:adynkFV}
\end{equation}

Eq. (\ref{equ:GnVV}) yields another adynkrafield
\begin{equation}
    \begin{split}
        \Hat{H}_{\un{a}}(x) ~=&~ h_{\un{a}}(x)
        ~+~ \ell\, \CMTred{\ytableaushort{\tinytwo}} \, h_{\b\a\Dot{\a}} (x)  
        ~+~ \Bar{\ell}\, \CMTred{\ytableaushort{\tinytwobar}} \,\Bar{h}_{\Dot{\b}\a\Dot{\a}}(x)   \\
        &~+~ \fracm{1}{2!}\,(\ell)^2\, h^{(2)}{}_{\un{a}}(x)  
        ~+~ \fracm{1}{2!}\,(\Bar{\ell})^2\, \Bar{h}^{(2)}{}_{\un{a}}(x)  
        ~+~ \ell\,\Bar{\ell}\,{\CMTB{\ydiagram{1}}} \,h_{\un{a}\un{b}}(x)  \\
        &~+~ \fracm{1}{2!}\,(\ell)^2\, \Bar{\ell}\,  \CMTred{\ytableaushort{\tinytwobar}}\,\Bar{\psi}_{\un{a} \, \Dot\b}(x) 
        ~+~ \fracm{1}{2!}\,(\Bar{\ell})^2\,\ell\, \CMTred{\ytableaushort{\tinytwo}}\, \psi_{\un{a} \, \b} (x) \\
        &~+~ \fracm{1}{2!2!}\,(\ell)^2\,(\Bar{\ell})^2\, A_{\un{a}}(x) ~~~.
    \end{split}
    \label{equ:adynkFHa}
\end{equation}
As a consistency check, tensoring the YTs on the right-hand side of Eq.~(\ref{equ:adynkFHa}) with ${\CMTB{\ydiagram{1}}}$ should reproduce the YTs appearing in the genome (\ref{equ:GnVV}). 
Indeed, tensoring with ${\CMTB{\ydiagram{1}}}$ yields
\begin{equation}
    \begin{split}
        {\CMTB{\ydiagram{1}}} \,\Hat{H}_{\un{a}}(x) ~=&~ {\CMTB{\ydiagram{1}}} \,h_{\un{a}}(x)
        ~+~ \ell\, \Big[\, {\CMTB{\ydiagram{1}}\CMTred{\ytableaushort{\tinytwo}}}_{\rm IR} \oplus \CMTred{\ytableaushort{\tinytwobar}}\, \Big] \, h_{\b\a\Dot{\a}} (x)  
        ~+~ \Bar{\ell}\, \Big[\, {\CMTB{\ydiagram{1}}\CMTred{\ytableaushort{\tinytwobar}}}_{\rm IR} \oplus \CMTred{\ytableaushort{\tinytwo}}\,\Big]  \,\Bar{h}_{\Dot{\b}\a\Dot{\a}}(x)   \\
        &~+~ \fracm{1}{2!}\,(\ell)^2\, {\CMTB{\ydiagram{1}}} \,h^{(2)}{}_{\un{a}}(x)  
        ~+~ \fracm{1}{2!}\,(\Bar{\ell})^2\,{\CMTB{\ydiagram{1}}} \, \Bar{h}^{(2)}{}_{\un{a}}(x)  \\
        & ~+~ \ell\,\Bar{\ell}\,\Big[\,{\CMTB{\ydiagram{2}}}_{\rm IR} \oplus \, {\CMTB{\ydiagram{1,1}}}_{\rm IR,+} \oplus \, {\CMTB{\ydiagram{1,1}}}_{\rm IR,-} \oplus \, \CMTB{\singlet}\,\Big] \,h_{\un{a}\un{b}}(x)  \\
        &~+~ \fracm{1}{2!}\,(\ell)^2\, \Bar{\ell}\,  \Big[\, {\CMTB{\ydiagram{1}}\CMTred{\ytableaushort{\tinytwobar}}}_{\rm IR} \oplus \CMTred{\ytableaushort{\tinytwo}}\,\Big] \,\Bar{\psi}_{\un{a} \, \Dot\b}(x) 
        ~+~ \fracm{1}{2!}\,(\Bar{\ell})^2\,\ell\, \Big[\, {\CMTB{\ydiagram{1}}\CMTred{\ytableaushort{\tinytwo}}}_{\rm IR} \oplus \CMTred{\ytableaushort{\tinytwobar}}\, \Big]\, \psi_{\un{a} \, \b} (x) \\
        &~+~ \fracm{1}{2!2!}\,(\ell)^2\,(\Bar{\ell})^2\,{\CMTB{\ydiagram{1}}} \, A_{\un{a}}(x) ~~~,
    \end{split}
    \label{equ:adynkFBoxHa}
\end{equation}
where the component fields $h_{\b\a\Dot{\a}} (x)$ and $\psi_{\un{a} \, \b} (x)$ correspond to ${\CMTB{\ydiagram{1}}}\otimes \CMTred{\ytableaushort{\tinytwobar}} ={\CMTB{\ydiagram{1}}\CMTred{\ytableaushort{\tinytwobar}}}_{\rm IR} \oplus \CMTred{\ytableaushort{\tinytwo}} $~, ${h}{}_{\un{a}}(x)$, ${h}^{(2)}{}_{\un{a}}(x)$, $\Bar{h}^{(2)}{}_{\un{a}}(x)$, and $A{}_{\un{a}}(x)$ corresponds to $\CMTB{\ydiagram{1}}$,
$\Bar{h}_{\Dot{\b}\a\Dot{\a}}(x)$ and $\Bar{\psi}_{\un{a} \, \Dot\b}(x) $ correspond to ${\CMTB{\ydiagram{1}}}\otimes \CMTred{\ytableaushort{\tinytwo}} ={\CMTB{\ydiagram{1}}\CMTred{\ytableaushort{\tinytwo}}}_{\rm IR} \oplus \CMTred{\ytableaushort{\tinytwobar}} $, 
and $h_{\un{a}\un{b}}(x)$ corresponds to ${\CMTB{\ydiagram{1}}}\otimes{\CMTB{\ydiagram{1}}} = {\CMTB{\ydiagram{2}}}_{\rm IR} \oplus \, {\CMTB{\ydiagram{1,1}}}_{\rm IR,+} \oplus \, {\CMTB{\ydiagram{1,1}}}_{\rm IR,-} \oplus \, \CMTB{\singlet}$. From this explicit computation, it is clear that the component fields and YTs appearing in Eq.~(\ref{equ:adynkFBoxHa}) are conjugate to one another. This is precisely what one expects, since ${\CMTB{\ydiagram{1}}} \,\Hat{H}_{\un{a}}(x)$ carries no free Lorentz indices.

Finally, Eq. (\ref{equ:GnVSSp}) yields an adynkrafield
\begin{equation}
    \begin{split}
        \Hat{L}_{\a}(x) ~=&~ l_{\a}(x)
        ~+~ \ell\,\CMTred{\ytableaushort{\tinytwo}}\, l(x)
        ~+~ \ell\,\CMTred{\ytableaushort{\tinytwo}}\, \tau_{\a\b}(x)
        ~+~ \Bar{\ell}\,\CMTred{\ytableaushort{\tinytwobar}}\, \xi_{\un a}(x)\\
        &~+~ \fracm{1}{2!}\,(\ell)^2\, \psi_{\a}(x)
        ~+~ \fracm{1}{2!}\,(\Bar{\ell})^2\,\epsilon_{\a}(x)
        ~+~ \ell\,\Bar{\ell}\,{\CMTB{\ydiagram{1}}}\, L^{(1)}{}_{\a\un b}(x)\\
        &~+~ \fracm{1}{2!}\,(\ell)^2\, \Bar{\ell}\, \CMTred{\ytableaushort{\tinytwobar}}\, L^{(2)}{}_{\un a}(x)
        ~+~ \fracm{1}{2!}\,(\Bar{\ell})^2\,\ell\,\CMTred{\ytableaushort{\tinytwo}}\, \sigma(x)
        ~+~ \fracm{1}{2!}\,(\Bar{\ell})^2\,\ell\,\CMTred{\ytableaushort{\tinytwo}}\, \omega_{\a\b}(x)\\
        &~+~ \fracm{1}{2!2!}\,(\ell)^2\,(\Bar{\ell})^2\, \eta_{\a}(x) ~~~,
    \end{split}
    \label{equ:adynkFLalpha}
\end{equation} 
where $L^{(1)}{}_{\a\un b}(x)$ corresponds to ${\CMTB{\ydiagram{1}}}\otimes \CMTred{\ytableaushort{\tinytwobar}} ={\CMTB{\ydiagram{1}}\CMTred{\ytableaushort{\tinytwobar}}}_{\rm IR} \oplus \CMTred{\ytableaushort{\tinytwo}} $. 
Also $\CMTred{\ytableaushort{\tinytwo}}\, \omega_{\a\b}(x) = -\fracm{i}{2}\,\CMTred{\ytableaushort{\tinytwo}}\,C^{\Dot\a\Dot\b}\, \omega^{(+)}_{\un a\un b}(x)$. 
If we impose the irreducible condition, $L^{(1)}{}_{\a\un b}(x)$ can be split into two irreducible component fields $\Psi{}_{\a\un b}(x)$ and $\Bar{\Psi}{}_{\Dot\a}(x)$. 

For completeness, the adynkrafield with a spinor index $\dot{\a}$ in the subscript reads 
\begin{equation}
    \begin{split}
        \Hat{\Bar{L}}_{\Dot\a}(x) ~=&~ \Bar{l}_{\Dot\a}(x)
        ~+~ \ell\,\CMTred{\ytableaushort{\tinytwo}}\, \bar{\xi}_{\un a}(x)
        ~+~ \Bar{\ell}\,\CMTred{\ytableaushort{\tinytwobar}}\,\bar{l}(x)
        ~+~ \Bar{\ell}\,\CMTred{\ytableaushort{\tinytwobar}}\,\bar{\tau}_{\Dot\a\Dot\b}(x) \\
        &~+~ \fracm{1}{2!}\,(\ell)^2\, \bar{\epsilon}_{\Dot\a}(x)
        ~+~ \fracm{1}{2!}\,(\Bar{\ell})^2\, \bar{\psi}_{\Dot\a}(x)
        ~+~ \ell\,\Bar{\ell}\,{\CMTB{\ydiagram{1}}}\, \Bar{L}^{(1)}{}_{\Dot\a\un b}(x)\\
        &~+~ \fracm{1}{2!}\,(\ell)^2\, \Bar{\ell}\, \CMTred{\ytableaushort{\tinytwobar}}\, \Bar{\sigma}(x)
        ~+~ \fracm{1}{2!}\,(\ell)^2\, \Bar{\ell}\, \CMTred{\ytableaushort{\tinytwobar}}\, \Bar{\omega}_{\Dot\a\Dot\b} (x)
        ~+~ \fracm{1}{2!}\,(\Bar{\ell})^2\,\ell\,\CMTred{\ytableaushort{\tinytwo}}\,\Bar{L}^{(2)}{}_{\un a}(x)\\
        &~+~ \fracm{1}{2!2!}\,(\ell)^2\,(\Bar{\ell})^2\, \Bar{\eta}_{\Dot \a}(x) ~~~,
    \end{split}
    \label{equ:adynkFLalphadot}
\end{equation}
where $\Bar{L}^{(1)}{}_{\Dot\a\un b}(x)$ corresponds to ${\CMTB{\ydiagram{1}}}\otimes \CMTred{\ytableaushort{\tinytwo}} ={\CMTB{\ydiagram{1}}\CMTred{\ytableaushort{\tinytwo}}}_{\rm IR} \oplus \CMTred{\ytableaushort{\tinytwobar}} $. 
Also $\CMTred{\ytableaushort{\tinytwobar}}\, \Bar{\omega}_{\Dot\a\Dot\b} (x) = -\fracm{i}{2}\,\CMTred{\ytableaushort{\tinytwobar}}\,C^{\a\b}\, \omega^{(-)}_{\un a\un b}(x)$.
If we impose the irreducible condition, $\Bar{L}^{(1)}{}_{\Dot\a\un b}(x)$ can be split into two irreducible component fields $\Bar{\Psi}{}_{\Dot\a\un b}(x)$ and $\Psi{}_{\a}(x)$.

\section{Conclusion}
\label{sec:CONs}

$~~~~$ In this work, we have undertaken the task of showing that the concept of an ``adynkrafield'' is consistent with the 
introduction of the Salam-Strathdee superspace definition in 4D, $\cal N$ = 1 of the supergravity prepotential as well as its
associated superfield description of the superconformal group.  The results in this paper also allow us to make deductions
about higher D theories.

This gives us more confidence in the correctness of 
the assertions in \cite{Cntg-11D}: that the eleven-dimensional scalar superfield $\cal U$ possesses a component-level 
symmetric graviton field at its sixteenth level, and that the eleven-dimensional spinor superfield ${\cal U}^{\a}$ possesses 
a component-level asymmetric graviton field at its seventeenth level.  The work in \cite{Cntg-11D} identified the
occurrence of the 11D YT's that correspond to the results shown in Eq. (\ref{eqn:Hag}).  

The properties of the 4D theory also allow us to make a conjecture about the reducibility of the 11D putative
prepotential scalar superfield $\cal U$. It is seen from Eq. (\ref{eqn:VCompDyn}) that there are two of the same
representations present at level 2.  We conjecture that such occurrences are a necessary, but not sufficient, 
feature indicating reducibility of any adynkra representation.  Looking at the results in \cite{Cntg-11D}, at the sixth
level there occur two
\begin{equation}
    \CMTB{[0,1,0,0,2]} ~~~
\end{equation}
 representations, which (according to our conjecture) imply that this is a reducible representation.  Thus, the next
 great challenge will be to marshal ideas about how to accomplish the complete reduction of  $\cal U$ into
 an irreducible representation of SUSY.

Currently, there is a collaboration \cite{DGHN} examining the conceptual basis for the proposal of adynkrafields 
on a more formal mathematical basis\footnote{Since the results were reported in the series of papers
\cite{Cntg-11D,Cntg-10D,Cntg-method,Cntg-lowD,Cntg-Weyl}, we have been informed that the \newline $~~~~~$ lower D adynkra 
genome examples may be interpreted as Hasse diagrams where the nodes correspond  \newline $~~~~~$
 to Schubert cycles.} that does not depend on Dynkin Labels. This work will be described in future work.
 This work may also provide a doorway to solving the remaining reducibility problem.

\vspace{.05in}
\begin{center}
\parbox{4in}{{\it ``
Start where you are. Use what you 
have. Do what you\\ $~~\,$ can.'' \\ ${~}$ 
 ${~}$ 
\\ ${~}$ }\,\,-\,\, Arthur Ashe $~~~~~~~~~$}
 \parbox{4in}{
 $~~$}  
 \end{center}
		
\noindent
{\bf Acknowledgments}\\[.1in] \indent

The research of S.J.G. is currently supported by the Clark Leadership Chair of Science endowment
at the University of Maryland.  The research of YH is supported by the Celestial Holography Initiative at the Perimeter Institute for Theoretical Physics and the Simons Collaboration on Celestial Holography. Research at the Perimeter Institute is supported by the Government of Canada through the Department of Innovation, Science and Industry Canada and by the Province of Ontario through the Ministry of Colleges and Universities.  Additional support for this
research was given in part by the endowment of the Ford Foundation Professorship of Physics at Brown 
University and the Brown Theoretical Physics Center. We very gratefully acknowledge calculations
undertaken by Dr.\ S.-N.\ Mak early in the development of this project and S.J.G. acknowledges T. H\" ubsch for
remarks regarding the reducibility conjecture.  Finally, we wish to acknowledge the use of ChatGPT 5 in the revision
process of this work.

\newpage
\appendix
\section{Dynkin Labels vs. Young Tableaux}

$~~~~~$ This appendix contains an introductory presentation on the complementary
perspectives of Klimyk's formula and the Young Tableau multiplication rule for their
operation on graphs of the form most familiar to physicists as Young Tableaux.  Given
two such graphs $V_\lambda$ and $V_\mu$,  Klimyk's formula and the Young Tableau 
multiplication rule are two different algorithms for computing the same 
representation-theoretic object \cite{FultonHarris,FultonYT}:
\begin{equation}
V_\lambda\otimes V_\mu
   =\bigoplus_{\nu}N_{\lambda\mu}^{\ \ \nu}V_\nu .
\end{equation}
They differ mainly in their language and range of applicability.

\subsection{Dynkin Labels \& Young Diagrams Encode $SU(N)$ Representation}

$~~~~~$ For $SU(N)$, let a Young diagram have row lengths
\begin{equation}
r_1\geq r_2\geq\cdots\geq r_{N-1}\geq 0.
\end{equation}
Its Dynkin Labels are
\begin{equation}
[a_1,a_2,\ldots ,a_{N-1}]
   =[r_1-r_2,\;r_2-r_3,\ldots ,r_{N-2}-r_{N-1},\;r_{N-1}].
\end{equation}

Conversely,
\begin{equation}
r_i=\sum_{j=i}^{N-1}a_j.
\end{equation}
Thus, for $SU(N)$, converting between Dynkin Labels and Young diagrams is simply a 
change of coordinates on the set of dominant highest weights \cite{Humphreys,Georgi}. For example, in $SU(3)$,
\begin{equation}
[p,q]\quad\longleftrightarrow\quad (r_1,r_2)=(p+q,q).
\end{equation}

Therefore, we have
\begin{equation}
[1,0]\longleftrightarrow \text{one box}, {~~~}
[2,0]\longleftrightarrow \text{two boxes in one row},{~~~}
[0,1]\longleftrightarrow \text{two boxes in one column}.
\end{equation}

\subsection{The Young Tableau Rule}

$~~~~~$ The Young-diagram or Littlewood--Richardson procedure multiplies two 
representations by adjoining the boxes of one diagram to the other, subject to the standard tableau conditions \cite{FultonYT,FultonHarris}:

\begin{itemize}
\item the Young-diagram shape conditions;
\item the row and column restrictions;
\item the Littlewood--Richardson lattice-word condition;
\item removal, for $SU(N)$, of complete columns of $N$ boxes \cite{Georgi}.
\end{itemize}

The allowed resulting diagrams, counted with their Littlewood--Richardson multiplicities, 
give the irreducible summands.  For example,
\begin{equation}
\text{one box}\otimes\text{one box}
   =
\text{two boxes in one row}\oplus\text{two boxes in one column},
\end{equation}
or in $SU(3)$ Dynkin Labels,
\begin{equation}
[1,0]\otimes[1,0]
   =[2,0]\oplus[0,1].
\end{equation}

\subsection{Klimyk's Formula}

$~~~~~$ Klimyk's formula works directly with weights and Weyl-group operations \cite{Klimyk1974,VilenkinKlimyk}. Let $\Lambda_1$ 
and $\Lambda_2$ be the highest weights of $V_{\Lambda_1}$ and $V_{\Lambda_2}$. For every 
weight $\lambda$ of $V_{\Lambda_1}$, with multiplicity $m_\lambda$, form

\begin{equation}
\mu=\lambda+\Lambda_2+\rho,
\end{equation}
where
\begin{equation}
\rho=\frac12\sum_{\alpha>0}\alpha
\end{equation}
is the Weyl vector. In Dynkin coordinates, $\rho$ has Labels

\begin{equation}
[1,1,\ldots ,1].
\end{equation}

One then:

\begin{enumerate}
\item reflects $\mu$ into the dominant Weyl chamber;
\item assigns the sign of the Weyl reflection;
\item discards terms lying on a Weyl-chamber wall;
\item subtracts $\rho$;
\item combines identical dominant weights with their signed multiplicities.
\end{enumerate}

Schematically,
\begin{equation}
V_{\Lambda_1}\otimes V_{\Lambda_2}
 =
\sum_{\lambda\in \mathrm{Wt}(V_{\Lambda_1})}
m_\lambda\,
\epsilon(\lambda+\Lambda_2+\rho)\,
V_{\{\lambda+\Lambda_2+\rho\}-\rho}.
\end{equation}
Here $\{\cdot\}$ means the Weyl-dominant representative. This weight-and-Weyl-group construction applies to both classical and exceptional semisimple Lie algebras \cite{Klimyk1974,VilenkinKlimyk}.

\subsection{Their Precise Relationship}

$~~~~~$ For type $A_{N-1}$ algebra of the Cartan Lie algebras, i.e. $SU(N)$). of Klimyk's formula 
and the Littlewood--Richardson rule produce exactly the same multiplicities.  The correspondence is:

\begin{center}
\begin{tabular}{ll}
\toprule
\textbf{Klimyk language} & \textbf{Young-tableau language} \\
\midrule
Highest weight & Young-diagram shape \\
Dynkin Labels & Differences of successive row lengths \\
Weights of an irrep & Semistandard tableau contents \\
Weyl reflections & Reordering rows and correcting illegal shapes \\
Weyl sign cancellation & Cancellation of inadmissible tableau contributions \\
Chamber-wall term & Vanishing or forbidden tableau configuration \\
Final dominant weight & Valid resulting Young diagram \\
Multiplicity & Number of Littlewood--Richardson tableaux \\
\bottomrule
\end{tabular}
\end{center}

The Young Tableau rule packages the type-$A$ tensor-product multiplicities into positive combinatorics: only valid 
tableaux survive \cite{FultonYT,FultonHarris}. Klimyk's formula allows intermediate terms with positive and negative signs; after 
Weyl reflection and cancellation, the surviving coefficients are the same nonnegative tensor-product 
multiplicities.

\subsection{Example: $SU(3)$, $6\otimes 3$}

$~~~~~$ The $6$ has Dynkin Label
\begin{equation}
[2,0]\longleftrightarrow\text{two boxes in one row},
\end{equation}
and the $3$ has
\begin{equation}
[1,0]\longleftrightarrow\text{one box}.
\end{equation}
The Young-diagram calculation gives
\begin{equation}
\text{two boxes in one row}\otimes\text{one box}
 =
\text{three boxes in one row}\oplus\text{the shape }(2,1).
\end{equation}
Converting back,
\begin{equation}
\text{three boxes in one row}\longleftrightarrow[3,0],\qquad
\text{the shape }(2,1)\longleftrightarrow[1,1].
\end{equation}

Therefore,
\begin{equation}
[2,0]\otimes[1,0]
   =[3,0]\oplus[1,1],
\end{equation}
or dimensionally,
\begin{equation}
6\otimes3=10\oplus8.
\end{equation}
This agrees with standard $SU(3)$ tensor-product tables \cite{Slansky}.
In Klimyk's calculation, several shifted weights occur, including terms that reflect to the same dominant 
weight with opposite signs. Those cancel, leaving precisely $[3,0]$ and $[1,1]$.

\subsection{Essential Distinction}

$~~~~~$ The Littlewood--Richardson/Young Tableau rule is especially natural and efficient for $SU(N)$ 
and $GL(N)$ \cite{FultonYT,Georgi}.  Klimyk's formula is more general: it applies uniformly to arbitrary semisimple Lie algebras, 
including groups such as \cite{Klimyk1974,VilenkinKlimyk}
\begin{equation}
SO(N),\quad Sp(2N),\quad G_2,\quad F_4,\quad E_6,\quad E_7,\quad E_8,
\end{equation}
where ordinary Young diagrams alone do not provide a complete universal multiplication rule.  Thus, 
the Young-tableau rule may be regarded as the particularly elegant type-$A$ combinatorial realization 
of the general highest-weight and Weyl-group mechanism expressed by Klimyk's formula.

$$~~$$

\end{document}